\documentclass[nofootinbib,eqsecnum,tightenlines,superscriptaddress,10pt]{revtex4}

\usepackage{graphicx}
\usepackage{wrapfig,lipsum}
\usepackage{amsmath,amssymb,amsfonts,amsthm,stmaryrd,mathtools,bm}
\usepackage{mathrsfs}
\usepackage{color}
\usepackage{multirow,bigdelim}
\usepackage{hyperref}
\usepackage{dsfont}
\usepackage{booktabs}
\allowdisplaybreaks[0]

\usepackage{pstricks}
\usepackage{tikz}
\usepackage{ulem}

\def\be#1\ee{\begin{align}#1\end{align}}

\def\ba{\begin{eqnarray}}
\def\ea{\end{eqnarray}}
\def\nn{\nonumber}
\def\q{\quad}

\begin{document}

\title{ From  spin foams to area metric dynamics to gravitons}
%\title{ Area metric and  graviton dynamics from the continuum limit of effective spin foams}

\author{Bianca Dittrich}
\affiliation{Perimeter Institute, 31 Caroline Street North, Waterloo, ON, N2L 2Y5, CAN}
\affiliation{Institute for Mathematics, Astrophysics and Particle Physics,
Radboud University, Heyendaalseweg 135, 6525 AJ Nijmegen, The Netherlands}
\author{Athanasios Kogios}
\affiliation{Perimeter Institute, 31 Caroline Street North, Waterloo, ON, N2L 2Y5, CAN}
\affiliation{Department of Physics and  Astronomy, University of Waterloo, 200 University Avenue West, Waterloo, ON, N2L 3G1, Canada}

\begin{abstract}
Although spin foams arose as quantizations of the length metric degrees of freedom, the quantum configuration space is rather based on areas as more fundamental variables. This is also highlighted by the semi-classical limit of four-dimensional spin foam models, which is described by the Area Regge action.

Despite its central importance to spin foams the dynamics encoded by the Area Regge action is only poorly understood, in particular in the continuum limit. We perform here a systematic investigation of the dynamics defined by the Area Regge action on a regular centrally subdivided hypercubical  lattice. This choice of lattice avoids many problems of the non-subdivided hypercubical lattice, for which the Area Regge action is singular. The regularity of the lattice allows to extract the continuum limit and its corrections, order by order in the lattice constant.

We show that, contrary to widespread expectations which arose from the so-called flatness problem of spin foams, the continuum limit of the Area Regge action does describe to leading order the same graviton dynamics as general relativity. The next-to-leading order correction to the effective action for the length metric is of second order in the lattice constant, and is given by a quadratic term in the Weyl curvature tensor.  
This correction can be understood to originate from an underlying dynamics of area metrics.

This suggests that the continuum limit of spin foam dynamics does lead to massless gravitons, and that the leading order quantum corrections can be understood to emerge from a generalization of the configuration space from length to area metrics.

\end{abstract}

\maketitle

%\tableofcontents

\section{Introduction}

Quantum gravity can be understood as the task to quantize (semi-) Riemannian geometry and to endow the resulting quantum geometry with a dynamics that leads to general relativity in the classical limit. But a number of approaches  suggests that more general forms of geometry might be better suited for a description of quantum gravity.  In particular in four dimensions, areas appear as more fundamental, due to their natural symplectic relationship to curvature angles \cite{BFCG2,EffSF1}. In (four-dimensional) loop quantum gravity  and spin foams \cite{LQG,Perez}, areas provide the fundamental geometrical degrees of freedom, and the corresponding quantum operators come with a discrete spectrum.  Areas also play a crucial role for the reconstruction of geometry from entanglement in holographic approaches to quantum gravity \cite{RyuTakayanagi}, as well as for black hole entropy counting arguments \cite{BHCounting}.  Area metrics have been also argued for as a tool to encode the phenomenology of  string theory \cite{Schuller1,Schuller2}.  

The Area Regge action provides a dynamics for a class of discrete geometries, in which areas are associated to triangulations. The Area Regge action is central in the dynamics of spin foams \cite{Perez}, as it appears in their semi-classical limit \cite{SFLimit} and provides the action for the recently introduced effective spin foam models \cite{EffSF1,EffSF2,EffSF3}. 

Despite this importance of the Area Regge action in quantum gravity, its dynamics has been poorly understood so far. It has been widely thought that the Area Regge action cannot lead to general relativity \cite{AreaRegge}. Indeed, the equations of motion seem to demand, that curvature is vanishing. This conclusion was based on a seemingly obvious but misguided identification of the deficit angle in Area Regge calculus with the deficit angle in Length Regge calculus \cite{Regge}. The latter  is a discretization of general relativity. Whereas the deficit angle in Length Regge calculus is indeed a measure for curvature associated to a Levi-Civita connection, the deficit angle in Area Regge calculus does measure rather a mixture of curvature and torsion. 

These misunderstandings  have  also contributed to the  so-called flatness problem for spin foams \cite{flatness,EffSF1,EffSF2}, which states the expectation that the semi-classical limit of spin foams suppresses curvature. There is more and more evidence \cite{MuxinSmallBI,EffSF1,EffSF2,EffSF3,ComplexSP} that this issue can be circumvented. The arguments and supporting examples in these works \cite{MuxinSmallBI,EffSF1,EffSF2,ComplexSP} are however based on triangulations with a small number of building blocks. They require furthermore  the implementation of area-length constraints, which force the areas to arise from a consistent length assignment.  With a strong implementation of these constraints the Area Regge action turns into the Length Regge action \cite{EffSF3}.  But, within the Hilbert space structure of loop quantum gravity, a strong imposition of these constraints is not possible, as these are second class \cite{DittrichRyan,EffSF1}. Whereas the Barrett-Crane spin foam model \cite{BC} does not implement these constraints,  the more recent spin foam models \cite{EPRL-FK, EffSF1,EffSF2} do impose the constraints weakly, with the Barbero-Immirzi parameter controlling  how much the constraints are allowed to fluctuate \cite{EffSF1}.  The works \cite{MuxinSmallBI,EffSF1,EffSF2,ComplexSP} therefore also suggest a bound on the Barbero-Immirzi parameter.

The recent work \cite{ARE1}, as well as this current work, do provide a completely independent mechanism for regaining propagating curvature degrees of freedom, i.e. gravitons from spin foams. This mechanism applies to triangulations with a large number of building blocks, and does not even require an implementation of area-length constraints and therefore also not a bound on the Barbero-Immirzi parameter. As the Area Regge action does appear in the semi-classical limit of spin foams and does provide the basis for the effective spin foam models, the results here showcase how spin foam dynamics can lead to propagating gravitons in the continuum limit.\footnote{The semiclassical limit for spin foams, which is described by the Area Regge action, can be equated with the limit of `large' areas in a given triangulation. But the correction terms are already extremely small, if the areas reach around a hundred Planck units. One can thus assume an average lattice constant of around 10 Planck length. Considering the continuum limit for the Area Regge action on the lattice means that we consider a limit where the wavelength of gravitons is much larger than this lattice constant. } We will moreover provide correction terms for the graviton dynamics, which arise from the fact that in spin foams areas are more fundamental than lengths. The correction terms can be also understood to arise from a dynamics defined for area metrics. 

To allow the consideration of triangulations with a large number of building blocks, we will consider on the one hand a regular lattice, and on the other hand apply an expansion of the configurations around a flat background field. This will allow us to show that (linearized) general relativity does emerge from (linearized) Area Regge calculus, at leading order in the lattice constant. The crucial mechanism\footnote{This mechanism has been predicted in the review \cite{ReggeWilliams}, but apparently computers and computational techniques have not been sufficiently advanced at the time for a complete analysis.} for this is the following: although Area Regge calculus features far more degrees of freedom than Length Regge calculus, most of these degrees of freedom will turn out to be massive, with the mass scaling with the inverse lattice constant. In fact, both Length Regge calculus and Area Regge calculus feature the same number of massless degrees of freedom, namely 10 per lattice vertex, which allow the reconstruction of the (length) metric.

On the other hand, one would expect from the fact that Area Regge calculus has much more degrees of freedom than e.g. Length Regge calculus, deviations from general relativity and thus phenomenological implications. Indeed, \cite{ARE1}, using the standard triangulation of the hypercubic lattice, computed the effect of these additional degrees of freedom onto the dynamics of the length metric. To leading order in the lattice constant, this effect can be summarized by a contribution to the action, which is of fourth order in the lattice constant (compared to zeroth order for the linearized Einstein-Hilbert term), quadratic in curvature and of sixth order in derivatives. 

Here we will find however, that this statement does depend on a certain feature of the standard triangulation of the hypercubic lattice: The Area Regge action is singular on this triangulation (or rather on the most regular background geometry for this triangulation). The work \cite{ARE1} considered therefore a tilting of this lattice, and then studied a limit where this tilting goes to zero. This limit does impose so-called hypercubical constraints, which already suppress almost half of the degrees of freedom. One could therefore ask, whether one will  find to leading order linearized general relativity also on lattices, where such hypercubical constraints are not encountered. 

Here we will answer this question to the affirmative. We will consider a slightly more involved triangulation, namely one based on a hypercubic lattice where every hypercube is subdivided by a vertex placed in its centre. Despite featuring double as many variables per vertex as the standard triangulation of the hypercubic lattice, a number of issues simplify as compared to \cite{ARE1}.  In particular, the Area Regge action is not singular on this triangulation, and correspondingly there are no hypercubical constraints.

The most crucial difference to the standard triangulation is that Area Regge calculus does now feature a sector, which can be interpreted in terms of area metrics. (This sector is suppressed by the hypercubical constraints in the standard triangulation.) In this sense the Area Regge action does provide also an action for area metrics. 

These area metric degrees of freedom provide also the lowest order correction to the graviton action at leading order. This correction is now of second order in the lattice constant, and can be described by an additional term added to the Einstein Hilbert action, which is a quadratic contraction of the Weyl curvature tensor. We will leave the study of the phenomenological implications of this term for future work.

On the more technical side, in this work we will  further develope a range of techniques, that facilitate the analysis of Length and Area Regge actions, as well as effective spin foam actions \cite{EffSF1,EffSF3}, on large lattices.  To showcase these techniques we will provide a Mathematica notebook \cite{LinkNB}, with which the calculations in this paper can be explicitly followed. The perturbative analysis of Regge actions started with the pioneering work \cite{RocekWilliams}.  In recent years these techniques have shown to be immensely useful for  the exploration of holography and one-loop corrections \cite{Hol3D4D,PImeasure1} and renormalization \cite{Improved,BahrDittrichHe}  in discrete gravity. We expect that these techniques will be essential to understand the dynamics of spin foams and related approaches in the continuum limit \cite{ContLimit}.

~\\
This paper can be read independently from \cite{ARE1}. The results presented here rely on calculations for which we used the symbolic computation program Mathematica. We will present all the necessary steps for this computation in the main text and in the appendices, but, apart from the final result, will avoid stating the rather large matrices, appearing at the various stages of the calculation. Readers interested in following and verifying the calculations in more detail are refereed to the Mathematica notebook \cite{LinkNB} associated to this paper. 

This paper is structured as follows: The next Section \ref{Sec:HC} considers kinematical aspects related to the Area Regge action.  We will give the necessary background on Area Regge calculus (Section \ref{RAC})  and construct the expansion of the Area Regge action on the centrally subdivided lattice in Sections \ref{CSHCL} to \ref{SecFtrafo}. We then consider a set of variable transformations in Sections \ref{Sec:separation} and \ref{Sec:Length}. These partition the area variables into dynamically distinguished sets, including the  length metric, as well as to the area metric fluctuations. 

Section \ref{AccEff} will then analyze the effective dynamics for the length metric, which is induced by the Area Regge action. This effective action is determined by integrating out all additional degrees of freedom from the Area Regge action in Section \ref{Sec:EffH}. Different types of degrees of freedom showcase a different scaling behavior in the Area Regge action, see Section \ref{Sec:Scaling}. This allows us to isolate the two leading terms in the effective action, which arise from the length and  area metric degrees of freedom, respectively. We compute these terms in Section \ref{LengthBlock} and \ref{Sec:Low}. In Section \ref{Sec:Weyl} we provide an astonishingly straightforward geometric interpretation of the correction term induced by the area metric.  

We close with a summary and outlook in Section \ref{Disc}. This includes a comparison of the dynamics of Area Regge calculus on  the centrally subdivided hypercubic lattice with the one on the standard triangulation of the hypercubic lattice. 

The Appendices \ref{AppA} to \ref{AppE} include a number of techniques needed for the analysis of  Area Regge calculus on regular lattices, as well as the definition of  spin projections on the hypercubic lattice.

\section{The linearized Area Regge action on the centrally subdivided hyper-cubical lattice}\label{Sec:HC}

In this section we will provide a short introduction to the Area Regge action (Section \ref{RAC}), and the necessary definitions and techniques to expand the Area Regge action on a regular lattice (Section \ref{CSHCL} to \ref{SecFtrafo}). 

We will then partition the area variables on our lattice into sets which are either geometrically or dynamically distinguished, in Sections \ref{Sec:separation} and \ref{Sec:Length}. This will include a partitioning into a Plus and Minus sector in Section \ref{Sec:separation}, with the Plus sector leading eventually to the area metric variables.  The area metric variables contain as a subset the length metric variables, discussed in Section \ref{Sec:Length}.

\subsection{The Regge action}\label{RAC}

The Length Regge action \cite{Regge} provides an approximation to the Einstein-Hilbert action on a triangulated manifold. The variables are given by the lengths associated to the edges of the triangulation. Equipped with these length data  the triangulation can be understood as a piecewise linear and piecewise flat manifold. The Length Regge action for a four-dimensional triangulation is then given by  
\ba\label{Eq:LR}
S_{\rm LR}\, =\,\frac{1}{\kappa} \sum_t A_t(L_e) \, \epsilon_t(L_e) \q ,
\ea
where $A_t(L_e)$ denotes the area of the triangle $t$ and 
\ba\label{defa1}
\epsilon^L_t(L_e)\,=\, 2\pi -\sum_{\sigma \supset t} \theta_t^\sigma(L_e)
\ea
is the deficit or curvature angle. $\theta_t^\sigma$ is the dihedral angle between two tetrahedra sharing the triangle $t$ in the four-simplex $\sigma$, and we sum in (\ref{defa1}) over all four-simplices sharing the triangle $t$.

The deficit angle $\epsilon_t(L_e)$  in Length Regge calculus measures the rotation that a vector undergoes, if parallel transported around a small loop in the plane orthogonal to the triangle $t$. For the definition of (Levi-Civita) parallel transport one uses the fact that two 4-simplices sharing a tetrahedron, and with a consistent length assignment, can be embedded isometrically into flat four-dimensional space.  Thus one can pull-back the Levi-Civita parallel transport of the flat four-dimensional space, for each pair of four-simplices, to the triangulation. Parallel transporting from four-simplex to four-simplex around a triangle, one will nevertheless find a net-result, which is described by the deficit angle.

The variation of the Length Regge action with respect to the length variables leads to the equations of motion\footnote{The Schl\"afli identity $\sum_{t\subset \sigma} A_t \delta \theta_t^\sigma=0$, which holds for all variations  $\delta$ of the geometry of the four-simplex $\sigma$, ensures that terms, which involve variations of the deficit angles  $\epsilon_t$ cancel out.}
\ba
\sum_t \frac{\partial A_t}{\partial L_e} \epsilon^L_t(L_e)  = 0\q .
\ea
These constitute a discretization of the vacuum Einstein equations, and thus admit curved solutions where $\epsilon_t\neq 0$.

The deficit angle (\ref{defa1}) can be computed from the dihedral angles $\theta_t^\sigma(L_e)$, which are defined locally on the four-simplices. A given four-simplex $\sigma$ has 10 edges and 10 areas, and therefore 10 length variables and 10 area variables. Locally in configuration space\footnote{ The functions $L_e^\sigma(A_t)$ are multi-valued as area squares are quadratic functions of the length squares, and one thus encounters a number of root choices in the inversion process. The  first order formulation of Area Regge calculus \cite{ADH1} provides a formalism where this issue is circumvented, but leads to the same dynamics as the second order action (\ref{ARA1}).} one can express the lengths of the four-simplex as functions of the areas $L_e^\sigma(A_t)$. This allows to express the dihedral angles of a given simplex $\sigma$ in terms of the areas of this simplex $\theta_t^\sigma(A_t)=
\theta_t^\sigma(L_e^\sigma(A_t))$.

We can thus define a `deficit angle' in terms of areas 
\ba\label{defa2}
\epsilon^A_t(A_t)\,=\, 2\pi -\sum_{\sigma \supset t} \theta_t^\sigma(A_t) \q .
\ea
We would like to caution the reader, that `deficit angle' is quite a misnomer: whereas the deficit angle in Length Regge calculus does measure curvature in a well defined sense, the deficit  angle in Area Regge calculus measures rather a conglomerate of curvature and a shape mismatch between neighboring four-simplices \cite{AreaAngle,DittrichRyan, ADH1}. 

This shape mismatch arises for the following reason: although we have the same number of area and length variables on a given simplex, gluing two simplices together along a tetrahedron (identifying  the length variables on its 6 edges and the area variables on its 4 triangles), we do have 16 area variables but only 14 length variables. Indeed, generically, the area variables for the two four-simplices will induce different lengths for the edges of the shared tetrahedron. The two additional variables can be understood to describe  differences between the three-dimensional dihedral angles in the tetrahedron, as induced by the  area assignments to the two four-simplices \cite{ADH1}.

Given, that we can define deficit angles in terms of areas, we can also express the Regge action purely in term of areas
\ba\label{ARA1}
S_{\rm AR} =\frac{1}{\kappa} \sum_t A_t \, \epsilon^A_t(A_t) \q .
\ea
This defines Area Regge calculus \cite{AreaRegge,ADH1}. Area Regge calculus is based on a much more general  configuration space than Length Regge calculus: we have discussed above, that two neighboring four-simplices share a generically shape-mismatched tetrahedron, and are, contrary to the situation in Length Regge calculus, not anymore embeddable into flat space. 

Varying the Area Regge action (\ref{ARA1}) with respect to the areas%\footnote{Here one employs again the Schl\"afli identity, which ensures that terms with variations of the deficit angle vanish.}
one obtains the equations of motion
\ba\label{ARA2}
 \epsilon^A_t(A_t) \,=\,0 \q ,
\ea
which demand that the (Area Regge calculus) deficit angles vanish. Despite these seemingly simple equations of motion, there are propagating degrees of freedom: Using the techniques of \cite{DittrichHoehn1} one can count the propagating degrees of freedom in local time evolution steps, and there are many more\footnote{Here a propagating degree is meant to be opposed to a gauge degree of freedom or a constraint degree of freedom. For a time evolution step at a vertex adjacent to $n$ edges, Length Regge calculus has $(n-4)$ propagating degrees of freedom whereas Area Regge calculus has $(3n-10)$ propagating degrees of freedom.} than in Length Regge calculus \cite{ADH1}.

Using a regular lattice, we will also find here that we have many more  degrees of freedom in Area Regge calculus than in Length Regge calculus. But many of the Area Regge calculus degrees of freedom will be massive. In fact we will find that Area Regge calculus and Length Regge calculus have the same number of massless degrees of freedom. In both cases these can be identified with the length metric.

\subsection{The centrally subdivided hypercubical lattice}\label{CSHCL}

Here we will consider a background geometry given by flat Euclidean 4D space. The triangulation will be based on a regular hypercubic  lattice  with lattice vectors
\ba\label{lvectors}
 e_0= \lambda(0,0,0,1)\; ,\q e_1=\lambda (0,0,1,0)\; ,\q e_2=\lambda (0,1,0,0)\;,\q e_3=\lambda (1,0,0,0) \q ,
 \ea
 where $\lambda$ is the lattice constant.
 We apply a periodic identification, so that we obtain a 4-torus with $N^4$ hypercubes, where $N$ can be an arbitrarily large number.
 
Each hypercube will be triangulated in the same way.  To explain this triangulation, we will be using a binary notation as employed in \cite{Mara,RocekWilliams,ARE1}: Given the hypercube spanned by the vectors (\ref{lvectors}) we interpret the coordinates of the hypercube's vertices as binary numbers. These provide us with the labels for the lattice's vertices: $v_0\equiv \lambda(0,0,0,0)$, $v_1=\lambda(0,0,0,1), v_2=\lambda(0,0,1,0),  v_3=\lambda(0,0,1,1), v_4=(0,1,0,0)$ and so on. In the following, we will often refer to vertices by their labels $v_0=\{0\},v_1=\{1\}$, and to simplices or more general building blocks by their vertex sets, e.g. $\{0,3\}$ describes the edge between the vertices $\{0\}$ and $\{3\}$. We will order the labels for a given simplex from smallest to largest.  With hypercuboid at vertex $\{X\}$ we mean the hypercuboid whose vertex with the smallest numerical value is given by $\{X\}$. 

\begin{wrapfigure}{r}{5cm}
\includegraphics[width=4.5cm]{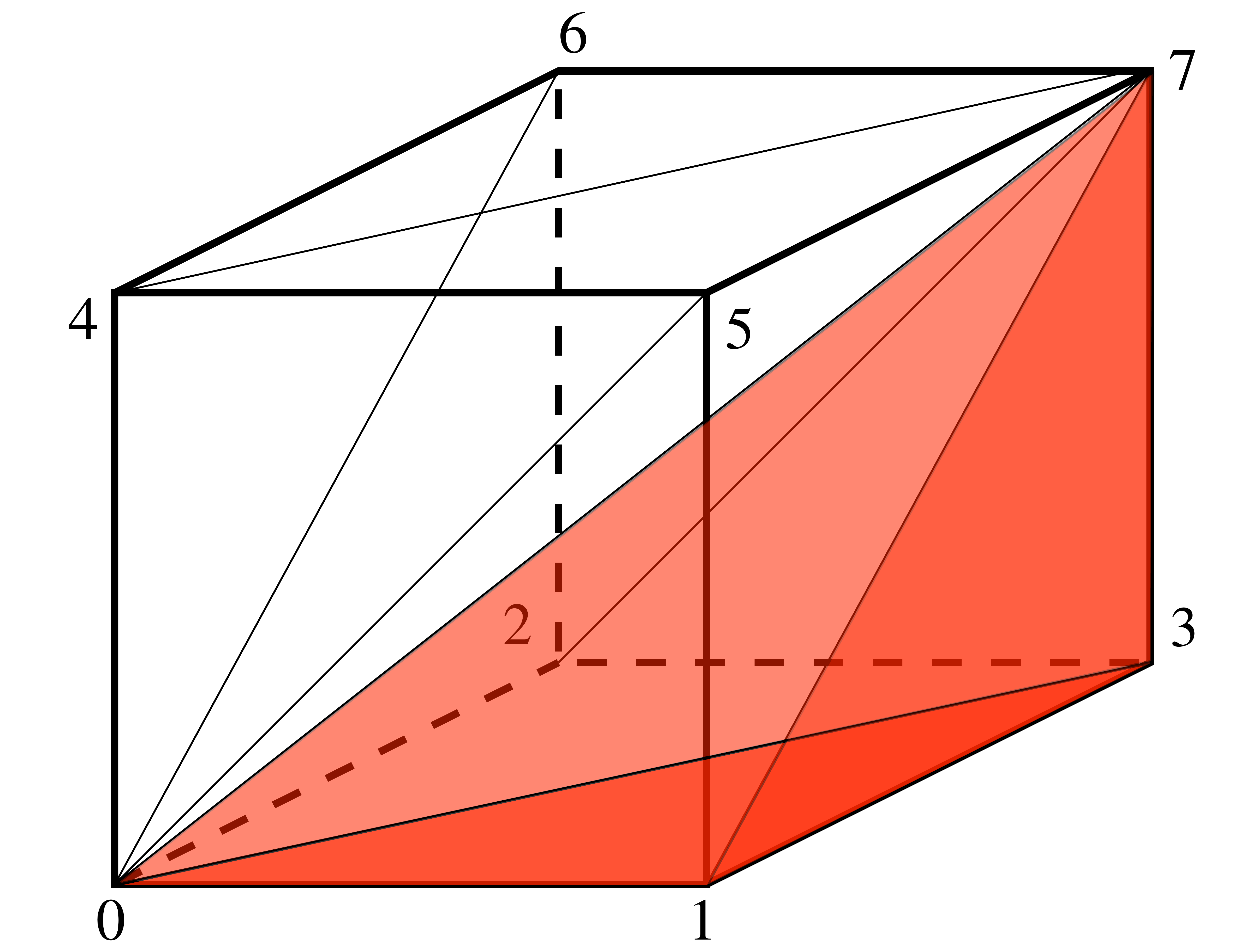}
\caption{Triangulation of the boundary cube $\{0,1,2,3,4,5,6,7\}$. The triangles of the tetrahedron $\{0,1,3,7\}$ are shown in different shades of red.}\label{3DCube}
\end{wrapfigure}

The work  \cite{ARE1}  considered the so-called standard triangulation \cite{Mara} of the hypercube, which was also used for the extraction of the continuum dynamics from Length Regge calculus in \cite{RocekWilliams}. Figure \ref{3DCube} shows the standard triangulation of the three-dimensional cube $\{0,1,2,3,4,5,6,7\}$, which consists of $6=3!$ tetrahedra of the form $\{0,2^i,2^i+2^j,7\}$. Here $i\neq j$ and $i,j$ can take values $i,j=0,1,2$. The standard triangulation of the hypercube $\{0,1,\ldots,15\}$ consists of $24=4!$ four-simplices of the form $\{0,2^i,2^i+2^j,2^i+2^j+2^k,15\}$, where $i,j,k$ are pairwise different and can take values $0,1,2$ or $3$. All of these four-simplices are isometric in the embedding into flat space defined by the lattice vectors (\ref{lvectors}) and contain the hyperdiagonal edge $\{0,15\}$. The hyperdiagonal edge is the hypothenuse for all the triangles it is shared with. Thus the derivative of the triangle areas with respect to the length of the hyperdiagonal, evaluated on the background geometry, vanishes. This leads to a singular structure for the expansion of the Area Regge action, which requires the inverse of the area-length derivatives.  The work \cite{ARE1} dealt with this issue by introducing a tilting with a parameter $s$ and considered the limit $s\rightarrow 0$ at the end of the calculation. The singular structure of the  Area Regge action induces so-called hypercubical constraints, that suppress a part of the area degrees of freedom in the $s\rightarrow 0$ limit.

Here we are looking for a triangulation of the hypercubic lattice, which does not lead to hypercubical constraints. In the standard triangulation it is the presence of the hyperdiagonal which leads to these constraints. One might therefore hope that avoiding triangulations which include the  hyperdiagonal, will also avoid hypercubical constraints. However the minimal triangulation of the hypercube \cite{Mara}, which consists of 16 four-simplices and does not feature the hyperdiagonal,  does nevertheless lead to hypercubical constraints.

An alternative option to avoid the hyperdiagonal is to use a triangulation of the hypercube with more four-simplices than the standard one. To construct such a triangulation for the hypercube, we choose to have the same triangulation of its three-dimensional boundary as in the standard triangulation of the  hypercube. That is the triangulation of each of the eight cubes is given by the standard triangulation for a three-dimensional cube \cite{Mara},  which consists of 6 tetrahedra, see Figure \ref{3DCube}.

Each of the $8\times 6=48$ tetrahedra in the boundary of the hypercube is then connected to a vertex in its centre (given by the coordinates $m_0=(\tfrac{1}{2},\tfrac{1}{2},\tfrac{1}{2},\tfrac{1}{2})$) for the hypercube at $\{0\}$), leading to 48 four-simplices. This defines our triangulation of the hypercube. By construction, it has the same three-dimensional boundary as the standard triangulation of the hypercube. This also means that we can glue hypercubes triangulated in this way directly to each other, as the triangulations of parallel boundary cubes agree with each other. (This is different for the minimal triangulation, where the orientations of (body-) diagonals differs between the parallel cubes of a given hypercube. Note also, that introducing an inner vertex for the minimally triangulated hypercube, in the same way as for the standard triangulation, does still lead to hypercubical constraints.)

This triangulation of the hypercube featuring a central vertex consists of 48  four-simplices,  48 tetrahedra in the boundary and 96 tetrahedra in the bulk, 96 triangles in the boundary and 64 triangles in the bulk, 64 edges in the boundary and 16 edges in the bulk. 

The  four-simplices are all isometric to each other in the background geometry.  The  edge length squares in the background geometry  are given by
\ba
{\bf L}^2=\lambda^2 \times (1,2,3,1,1,2,1,1,1,1) \q .
\ea
where,  for the edges of the four-simplex $\{ 0, 1, 3, 7, m_0\}$, we used the ordering:
$$\{\{0, 1\}, \{0, 3\}, \{0, 7\}, \{0, m_0\}, \{1, 3\}, \{1, 7\}, \{1, m_0\}, \{3, 7\}, \{3,
   m_0\}, \{7, m_0\}\}.$$

\subsection{The Area Regge Hessians for the four-simplices}

We can compute the linearized Area Regge action for the lattice by computing the Area Regge Hessian for each four-simplex, and then sum the contributions coming from all four-simplices.

We will use fluctuation variables  $\alpha_t$ associated to a triangle $t$ defined by $A^2_t= {\bf A}_t^2 + \alpha_t$, where ${\bf A}_t^2$ is the background value of the squared area for the triangle $t$.  On  a background with vanishing deficit angles  the quadratic part of the action is given by 
\ba\label{SAR2}
 S^{(2)}_{AR}=\tfrac{1}{2\kappa} \sum_{t,t'} \alpha_t  H_{tt'}\alpha_{t'}  \q \text{with} \q H_{tt'} = - \sum_\sigma \sum_{e\in \sigma}\frac{1}{2A_t}  \frac{\partial \theta_t}{\partial L^2_e}    (J^\sigma)^{-1}_{et'} \q 
\ea
with $\kappa=8\pi G/c^3$ and where $J^\sigma_{te}=\partial A_t^2/\partial L^2_e$ is the Area-Length Jacobian, formed from the lengths square derivatives of the area squares, and associated to the simplex $\sigma$. Importantly, the Hessian is given by a sum over the four-simplices. We can thus compute the Hessian separately for each simplex, and then sum the contributions over all four-simplices in the lattice.

The derivatives $\frac{\partial \theta_t}{\partial L^2_e}$ can be computed for a general four-simplex \cite{DittrichFreidelSpeziale} and evaluated for the background geometry of the four-simplices.  The same holds for the matrix of area square -- length square derivatives. For the lattice at hand, these Jacobians will be invertible, that is not feature null-vectors. We will therefore avoid the appearance of hyper-cubical constraints \cite{ARE1}.

The Hessian $H^\sigma$ for a four-simplex in our triangulation can be evaluated to 
\ba
H^\sigma={\tiny -\frac{1}{3\lambda^6}
\left(
\begin{array}{cccccccccc}
 0 & 0 & 0 & 3 & -12 & 12 & -6 & 12 & -6 & 0 \\
 0 & 3 & -6 & -6 & 12 & -6 & 3 & -6 & 0 & 6 \\
 0 & -6 & 16 & 6 & -12 & 0 & 0 & 0 & 0 & 0 \\
 3 & -6 & 6 & 3 & 0 & -6 & 0 & -6 & 12 & -6 \\
 -12 & 12 & -12 & 0 & -6 & 12 & -6 & 24 & -18 & 0 \\
 12 & -6 & 0 & -6 & 12 & -8 & 12 & -24 & 12 & 0 \\
 -6 & 3 & 0 & 0 & -6 & 12 & 0 & 12 & -12 & 0 \\
 12 & -6 & 0 & -6 & 24 & -24 & 12 & -40 & 24 & 0 \\
 -6 & 0 & 0 & 12 & -18 & 12 & -12 & 24 & -6 & -12 \\
 0 & 6 & 0 & -6 & 0 & 0 & 0 & 0 & -12 & 16 \\
\end{array}
\right)
}
\ea
To state the labelling of the rows and columns of $H^\sigma$, we pick the simplex $\sigma= \{ 0, 1, 3, 7, m_0\}$. We then use the following ordering for its triangles: $$\{\{0, 1, 3\}, \{0, 1, 7\}, \{0, 1, m_0\}, \{0, 3, 7\}, \{0, 3, m_0\}, \{0, 7, 
  16\}, \{1, 3, 7\}, \{1, 3, 16\}, \{1, 7, m_0\}, \{3, 7, m_0\}\}.$$
This Hessian $H^\sigma$ has one null vector $n_\sigma$ given by 
\ba
 (n_\sigma)^T={\bf A}^2=\lambda^4\times (\tfrac{1}{4}, \tfrac{1}{2},\tfrac{3}{16}, \tfrac{1}{2}, \tfrac{1}{4}, \tfrac{3}{16}, \tfrac{1}{4}, \tfrac{3}{16}, \tfrac{1}{4}, \tfrac{3}{16}) \q ,
 \ea
  where we denote with $n^T$ the transpose of a vector $n$. The entries coincide with the area squares of the triangles of the simplex, in the background geometry. The null vector $n_\sigma$ describes therefore a scaling symmetry.  This global scaling symmetry is also present for the non-perturbative Area Regge action. But the scaling symmetries for the Hessians of the four-simplices lead also to a local symmetry: \cite{DittrichFreidelSpeziale} shows that the scaling symmetries for the simplex Hessians are the building blocks for the (linearized) diffeomorphism symmetries of the lattice Hessian.

The Hessian for the Area-Regge action can be also calculated by an alternative method, which is detailed in \cite{ADH1}. It uses derivatives of the determinant of the Angle Gram matrix associated to the four-simplex with respect to its dihedral angles, and derives from the first order formulation of Area Regge Calculus \cite{ADH1}.

\subsection{The Area Regge Hessian for the hypercube}

The Hessians for the four-simplices $\sigma$ in a given hypercube ${\bf C}$
\ba
H^{\sigma}_{tt'} &=&- \sum_\sigma \sum_{e\in \sigma}\frac{1}{2A_t}  \frac{\partial \theta_t}{\partial L^2_e}    (J^\sigma)^{-1}_{et'} 
\ea
can be summed and give the Hessian for this hypercube:
$
H^{\bf C}_{tt'}= \sum_{\sigma \in {\bf C}}  H^{\sigma}_{tt'}\, .
$
The associated quadratic action can be split into bulk-bulk, boundary-boundary and bulk-boundary contributions:
\ba
S^{\bf C}&=&  \tfrac{1}{2\kappa} \bigg(  \sum_{t^{}_b,t'_b \in {\cal B}({\bf C})}\!\!\!
 \alpha_{t^{}_b}  H^{{\bf C}_{bb}}_{t^{}_b t'_b} \alpha_{t'_b} \,+\, 
  \sum_{t^{}_i,t'_i \in {\cal I}({\bf C})}\!\!\! \alpha_{t^{}_i}  H^{{\bf C}_{ii}}_{t^{}_i t'_i} \alpha_{t'_i} \,+\, %\nn\\
 % &&\q
 \!\!\!\!\!\!
 \sum_{t^{}_b \in {\cal B}({\bf C}) ,t'_i \in {\cal I}({\bf C})} \!\!\!\!\!\!\! \alpha_{t^{}_b}  H^{{\bf C}_{bi}}_{t^{}_b t'_i} \alpha_{t'_i} \,+\, 
 \!\!\!\!\!\!
  \sum_{t^{}_i \in {\cal I}({\bf C}) ,t'_b \in {\cal B}({\bf C})}\!\!\! \!\!\!\!\!\!\! \alpha_{t^{}_i}  H^{{\bf C}_{ib}}_{t^{}_i t'_b} \alpha_{t'_b}
\bigg) \, ,
\ea
where we denote with ${\cal B}({\bf C})$ the boundary and with  ${\cal I}({\bf C})$ the interior of the hypercube ${\bf C}$.

We have 96 boundary triangles and 64 bulk triangles, that is overall 160 variables for the Area Regge action associated to the hypercube. But we can reduce the number of variables to the 96 associated to the boundary triangles by integrating out the variables associated to the bulk triangles.

That is, we aim to define an effective action 
\ba
S^{\bf C'} &=& 
\tfrac{1}{2\kappa} 
\sum_{t^{}_b,t'_b \in {\cal B}({\bf C})}  \!\! \alpha_{t^{}_b}  H^{{\bf C'}}_{t^{}_b t'_b} \alpha_{t'_b}
 \q ,
\ea
 which only involves the variables associated to the boundary of the hypercube. The Hessian matrix describing the effective action is in principle given by
\ba\label{EffH1}
H^{{\bf C'}}_{t^{}_b t'_b} &=&H^{{\bf C}_{bb}}_{t^{}_b t'_b}  - \sum_{t^{}_i,t'_{i} \in {\cal I}({\bf C})} H^{{\bf C}_{bi}}_{t^{}_b t^{}_i}\,\, 
\left( H^{{\bf C}_{ii}}\right)^{-1}_{t^{}_i t'_i}\,\,
H^{{\bf C}_{ib}}_{t'_i t'_b}      \q .
\ea
But $H^{{\bf C}_{ii}}$ is not invertible: it has four null vectors  $n^j,j=0,\ldots, 3$ that correspond to the expected diffeomorphism symmetry \cite{RocekWilliams,DiffReview08,ADH1}, associated to the central vertex of the hypercube. By adding zero's for all entries labelled by boundary triangles ${t_b}$, we can extend these  four null vectors to null vectors of the full Hessian $H^{\bf C}$. That is the four null vectors are also null vectors for $H^{{\bf C}_{bi}}$. The issue of non-invertibility can therefore be dealt with by adding to $H^{{\bf C}_{ii}}$ a (gauge fixing) term
\ba
{\cal G}_{t^{}_i  t'_i} \,=\, \sum_{j=0,\ldots,3}  \beta_j  \,  (n^j)_{t^{}_i} (n^j)_{t'_i}
\ea
where $\beta_j$ are non-vanishing, but otherwise arbitrary gauge parameters. The matrix $H^{{\bf C}_{ii}} +{\cal G}$ is then invertible, and we can use this inverse to compute the Hessian (\ref{EffH1}) for the effective action. Because the $n^j$ are also null vectors for $H^{{\bf C}_{bi}}$ the effective Hessian will not depend on the gauge parameters $\beta_j$.

Note that this procedure of integrating out the bulk area squares can be also understood in terms of decoupling  a certain set of variables \cite{ARE1}. These variables are basically given by the variations of the action with respect to the bulk area squares. These  amount to (a rescaling of) the deficit angles of Area Regge calculus. Thus integrating out the bulk triangles, imposes that these deficit angles associated to triangles in the bulks of the hypercubes, vanish in average.

The Hessian $H^{{\bf C'}}$ for the effective action for the hypercube is a $96\times 96$ matrix, and still somewhat unwieldy. Going to the lattice and using a lattice Fourier transform, we can describe the dynamics in terms of a $36\times 36$ momentum dependent Hessian matrix, whose construction we will describe next.

\subsection{The (effective) Area Regge Hessian on the lattice and Fourier transform}\label{SecFtrafo}

To obtain the action for the entire lattice we can sum the contribution from each of its hypercubes:
\ba\label{LHess1}
(S')^{(2)}_{AR}&=& \sum_{\bf C} \sum_{t,t'}  \alpha_t  H^{{\bf C'}}_{tt'} \alpha_{t'} \,=\,   \sum_{\nu,\nu'}  \sum_{t@\nu,t'@\nu'} \alpha_t(\nu)  \,H_{tt'}(\nu,\nu') \,\alpha_{t'}(\nu') \q .
\ea
As we integrated out the bulk variables from each hypercube, we include in the first double sum in (\ref{LHess1}) only triangles $t,t'$ that are in the boundary of some hypercube. Such triangles are normal (in the background geometry) to either of the lattice vectors $e_j,j=0,\ldots,3$. 

In the second equation in (\ref{LHess1}) we just reorganized the sum. To this end we associated each triangle $t=\{ X,Y,Z\}$, where $X<Y<Z$, to the vertex $X$. We identified this vertex $X$ with its lattice coordinates $\nu\equiv (\nu_3,\nu_2,\nu_1,\nu_0)\in \mathbb{N}^4$. As explained in Section \ref{CSHCL}, $X$ arises from interpreting the $\nu$--coordinates as binary number $X=\sum_{j=0}^1 \nu_j2^j$.  The triangles associated to a vertex $X$ are all included in the boundary of the hypercube at $X$, but not all of the triangles in the hypercube's boundary are associated to the vertex $X$.    There are 36 triangles associated to a given vertex $\nu$ and thus an associated set of 36 variables $\{\alpha_t(\nu)\}_{t@\nu}$ for each vertex $\nu$.   

In the following we introduce a new label for the triangles $\tilde t$, which identifies the triangles associated to a given vertex $\nu$, and thus takes on 36 values. This new label $\tilde t$ can be obtained from quotienting the original labels $t\equiv (X,Y,Z)$ by an equivalence relation, given by $(X,Y,Z)\equiv (X',Y',Z')$ if $X=X'+z, Y=Y'+z$ and $Z=Z'+z$ for a $z\in \mathbb{Z}$.  We can now uniquely identify each triangle $t$ in the lattice, which is in the boundary of some hypercube, by the index pair $(\nu,\tilde t)$, and can thus label our variables as $\alpha_{\tilde t}(\nu)$. 

This allows us to define a lattice Fourier transform for the variables given by
\ba\label{FTrafo}
\alpha_{\tilde t}(\nu)=\frac{1}{\sqrt{N^4}}\sum_k \exp( \imath\,  \Lambda\, \sum_{j=0}^3 k_j\cdot \nu_j )\, \alpha_{\tilde t}(k) \q .
\ea
Here we introduced the momentum variables $k\equiv(k_3,k_2,k_1,k_0)$ with $k_i=2\pi K_i/(N \lambda)$ and $K_i=0,1,\ldots,N-1$ for $i=0,1,2,3$.
 We will set $\Lambda$ eventually equal to the background lattice constant $\lambda$. We will keep $\Lambda$ for now, as it allows us to count the powers of the momenta $k$ in a given expression.  The continuum limit will be defined as $\lambda\rightarrow 0, N\rightarrow \infty$ so that $N   \lambda$ remains constant. 
 
This Fourier transform allows us to encode the Hessian as a $36\times 36$ momentum-dependent matrix $H_{tt'}(k')$, defined by
\ba
(S')^{(2)}_{AR}&=& \sum_{\nu,\nu'}  \sum_{\tilde t,\tilde t'} \alpha_{\tilde t}(\nu)  \,H_{\tilde t \tilde t'}(\nu,\nu') \,\alpha_{\tilde t'}(\nu') \,\,=\,\,\sum_{k,k'} \sum_{\tilde t, \tilde t'}\alpha_{\tilde t}(k) H_{\tilde t \tilde t'}(k')\, \delta^{(4)}_{-k,k'} \, \alpha_{\tilde t'}(k') \q ,
\ea
 where  $\delta^{(4)}_{-k,k'}=\delta_{-k_0,k'_0}\cdots \delta_{-k_3,k'_3}$.  For instance, a  term $\delta^{(4)}_{\nu,\nu'}$ in $H(\nu,\nu')$ Fourier transforms to $1$ in $H(k')$,  and a term $\delta^{(4)}_{\nu\pm\rho,\nu'}$ Fourier transforms to $(\omega_\rho)^{\pm1}:=\exp(\pm\imath \Lambda \sum_{j=0}^3 k'_j \rho_j)$.  Thus, the matrix entries of $H(k')$ are functions of the  $\omega_\rho$ and thus the $k'_j$. We can expand the $(\omega_\rho)^{\pm 1}$ factors in $\Lambda$. This leads to an expansion of  $H(k')$ in positive powers of $\Lambda$. The coefficient of $\Lambda^n$ is a homogeneous polynomial of degree $n$ in the momenta $k'_j,j=0,\ldots 3$.
 
 The $\Lambda^0$-coefficient of $H_{\tilde t,\tilde t'}(k')$ can be understood to define a mass matrix for the variables $\alpha_{\tilde t}(k')$. This mass matrix scales with a homogeneous pre-factor $\lambda^{-6}$, indicating that all modes with a non-vanishing eigenvalue with respect to this mass matrix are getting very massive and their propagation will therefore be suppressed in the continuum limit.
 
 The mass matrix, given by the $\Lambda^0$-coefficient of $H_{\tilde t,\tilde t'}(k')$, has  26 non-vanishing eigenvalues and 10 null vectors. We will see that we can identify the 10 null vectors with the 10 degrees of freedom in the length metric.  To make this behavior more transparent, we will perform a splitting of the variables into various sets.  The corresponding blocks of the Hessian matrix will show different scaling behavior in $\Lambda$. But one can also argue for such a split based on the geometric interpretation of the variables.

\subsection{Transformation to Area metric variables}\label{Sec:separation}

Here we will define further variable transformations and then define four different sets of variables. We will see that these different sets of variables will lead to different scaling behavior of the associated Hessian blocks in $\Lambda$. A similar type of transformation has been already used for the standard triangulation of the hypercubical lattice in \cite{ARE1}. 

The first transformation is motivated by the hope to identify a sector that describes the dynamics of an area metric \cite{Schuller1,Schuller2}. An area metric measures the areas of parallelograms and the dihedral angles between these. But here we are using the area of triangles, not of parallelograms. 
We therefore define two types of variables: the first type $\alpha^+$ are given by sums of area square fluctuations associated to two triangles in the same plane, and the second type $\alpha^-$ as differences thereof. Whereas $\alpha^+$ measures the area of the quadrilateral formed by the two triangles, $\alpha^-$ provides one measure of how much this quadrilateral deviates from a parallelogram.

\begin{wrapfigure}{r}{6cm}
\hspace{0.5cm}
\includegraphics[width=4.5cm]{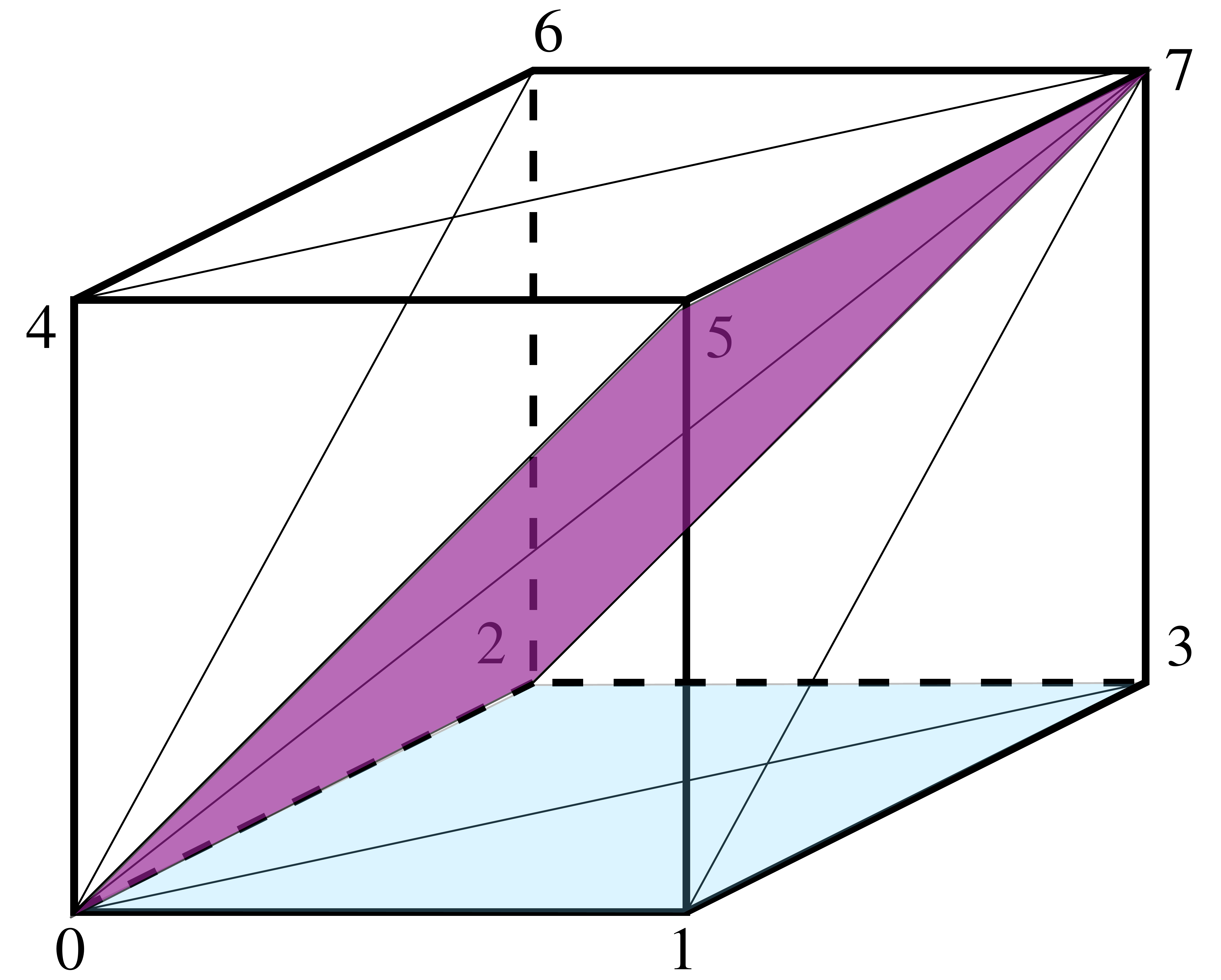}
\caption{A triangle pair $(1,2)\equiv (\{0,1,3\},\{0,2,3\})$ of the first type,  in blue, and a pair $(2,5)\equiv(\{0,2,7\},\{0,5,7\})$ of the second type, in purple. \label{PlanesPic}}       
\end{wrapfigure}

To detail this transformation we introduce the following notation for the triangles $\{Y,Y',Y''\}$ associated to a given vertex $\{Y\}$. To identify parallel triangles associated to different vertices we translate them all to $\{0\}$, and use the label $\{0,Y'-Y,Y''-Y\}$. We then have two types of triangles in the boundary of the hypercubes, which are illustrated in Figure \ref{PlanesPic}:

\begin{itemize}
\item Triangles in a plane spanned by two of the lattice vectors $\{e_0,\ldots, e_3\}$. These are of the form $\{0,2^i, 2^i+2^j\}$, where $i,j=0,\ldots,3$ and $i\neq j$. There are 12 such triangles. We can form 6 pairs of triangles in the same plane, these are given by $(\{0,2^i,2^i+2^j\},\{0,,2^j,2^i+2^j\})$, where now $i<j$. 
\item Triangles in a plane spanned by three of the lattice vectors $\{e_0,\ldots, e_3\}$. There are 24 such triangles, from which we can again define 12 pairs formed by triangles situated in the same plane. These pairs are given by $(\{0,2^i,2^i+2^j+2^k\},\{0,2^j+2^k,2^i+2^j+2^k\})$, where $i,j,k=0,\ldots 3$ and are mutually different. (Note that exchanging $j$ and $k$ defines the same pair of triangles.)
\end{itemize}

We define the variables $\alpha^+$ and $\alpha^-$ as sums and differences of the area square fluctuations, associated to these pairs of triangles:
\ba\label{trafo2}
\alpha^+_{(X,X')}&=&\frac{1}{\sqrt{2}} \left( \omega^{-1}_X \,\alpha_{\{0,X,X+X'\}}+  \omega^{-1}_{X'}  \,\alpha_{\{0,X',X+X'\}} \right)  \, , \nn\\
\alpha^-_{(X,X')}&=&\frac{1}{\sqrt{2}} \left(  \omega^{-1}_X \, \alpha_{\{0,X,X+X'\}} -   \omega^{-1}_{X'} \, \alpha_{\{0,X',X+X'\}} \right)  ,\;\;\;
\ea
where we consider pairs $(X,X')$ with $X=2^i$ and $X'=2^j$, in which case we assume $i<j$,  as well as pairs with $X=2^i$ and $X'=2^j+2^k$ for $i,j,k=0,\ldots,3$ and mutually different, but we do {\it not} assume that $i<j$ or $i<k$. %That is the edge $\{0,X\}$ is parallel to a basic lattice vector, whereas $\{0,X'\}$ is either parallel to a basic lattice vector or to the sum of two basic lattice vectors.

The momentum factors $\omega_{X}$ with binary label $X=\sum_i b_i 2^i$ are given by $\omega_{X}= \exp(\imath \Lambda \sum_{j=0}^3 b_j k_3)$.  The $\omega$-factors define lattice shifts  and thus specify which two triangles in a given plane we consider, namely $\{-X,0,X'\}$ and $\{-X',0,X\}$ respectively. (These two triangles do not form a quadrilateral, but the resulting object can be considered to be centered at $\{0\}$.)  This choice of $\omega$-factors above does simplify the momentum dependence of the length square fluctuations expressed in terms of these new variables. We will see that this will allow us to define projectors which specify the area-lengths constraints, and do not contain a momentum dependence, if expressed in this new basis. (The same $\omega_X$ and $\omega_{X'}$ factors were used in \cite{ARE1}, where these led to projectors on the hypercubical constraints without a momentum dependence.)
Having matrices without a momentum dependence can hugely simplify the calculations, e.g. when having to invert a (large) matrix.

We can  write the transformation of variables ddefined in (\ref{trafo2}) in a  condensed notation as 
\ba
\alpha\,=\, {\mathbb P}\cdot \left(\begin{matrix} 
\alpha^+ \\
\alpha^-
\end{matrix}\right)\,=\,
\left(\begin{matrix}
P^+  &
 P^- 
\end{matrix}
\right)\cdot \left(\begin{matrix} 
\alpha^+ \\
\alpha^-
\end{matrix}\right)
\ea
where $P^{+},P^{-}$ denote $18\times 36$ (momentum dependent) matrices and ${\mathbb P}$ is a unitary matrix. 
Let us remark that there is an arbitrariness in the choice of signs for the differences $\alpha^-_{(XX')}$.  Hessian blocks and projectors involving the $\alpha^-$ variables might depend on this choice. But we will see that the $\alpha^-$ variables will not contribute to the leading and next to leading order in the effective action describing the dynamics of the length metric degrees of freedom.

We thus separate our variables $\{\alpha_{\tilde t}(k)\}$ into two sets $\{\alpha^+_{X,X'}\}$ and $\{\alpha^-_{X,X'}\}$. We will refer to these two sets as Plus and Minus sector respectively. We will see below that the Hessian blocks with respect to these two sectors exhibit different scaling behavior in $\Lambda$, that is these two sectors are distinguished by the dynamics. 

For the Plus sector we can now interpret the variables as area square fluctuations associated to planes and we can aim to define a transformation to variables which can be interpreted as area metric fluctuations.  To keep the same index structure for the Plus and Minus sector, we will apply the same transformation to the Minus sector. 

Let us first say a few words about the area metric \cite{Schuller1,Schuller2}. 
 In the same way as the length metric $g$ measures the lengths of vectors and angles between vectors, the area metric $G$ measures the areas of parallelograms (spanned by two vectors) and dihedral angles between these parallelograms.  We can denote the parallelograms spanned by two vectors $a$ and $b$ by the wedge product $a\wedge b$. We can define  6 parallelograms, which are spanned by pairs of basis (lattice) vectors, and which we can denote by indices $(ij)\equiv e_i\wedge e_j$ with $i<j$ and $i,j=0,\ldots,3$. For notational convenience we will also allow indices $(ij)$ with $j<i$ with the identification $e_i\wedge e_j=-e_j\wedge e_i$.  The area metric is a symmetric, bi-linear form  on the space of parallelograms and its components $G_{(ij)(km)}$ therefore define a symmetric matrix with respect to the double-indices.  The area metric has thus $21=6\times 5/2+6$ independent components.\footnote{One can furthermore demand  the cyclicity condition $G_ {(ij)(kl)}+G_ {(ik)(lj)}+G_ {(il)(jk)}=0$, as done in \cite{Schuller1}.  If all other (anti-) symmetry requirements are satisfied, one obtains only one further independent equation for the components, namely $G_{(01)(23)} + G_{(02)(31)}+ G_{(03)(12)}=0$. Thus, imposing cyclicity, the area metric has  20 independent components and the area metric at a fixed space-time point constitutes a 20-dimensional irreducible representation space for $\text{SL}(4,\mathbb{R})$.  This requirement is however dropped in other works on the area metric, e.g. \cite{Schuller2}.  Note that the area metric has the same set of algebraic symmetries as the Riemann tensor --- the cyclicity condition corresponds to the so-called algebraic Bianchi identity for the Riemann tensor.}
 
 We do however have only 18 variables in the Plus and Minus sector, respectively. Indeed, as we integrated out triangles in the bulk of the hypercubes, we are missing all entries of the Area metric which involve two planes that span the entire four-dimensional space. This accounts for the 3 missing independent components $G_{(01)(23)},G_{(02)(13)},G_{(03)(12)}$. We can nevertheless interpret our result in terms of the Area metric, if we assume likewise that the 3 missing components have been integrated out.

The transformation between the area metric components and the area squares of the basic parallelogram follows from the bi-linearity of the area metric
 \ba\label{AMAtrafo1}
A^2(e_i\wedge e_j)&=& G( e_{i} \wedge e_{j},  e_{i} \wedge e_{j})\q\q\q\q\q\q=\, G_{(ij)(ij)}  \q , \nn\\
 A^2(e_i\wedge (e_j+e_k))&=& G( e_{i} \wedge (e_{j}+e_{k}),  e_{i} \wedge  (e_{j}+e_{k}))\,=\, G_{(ij)(ij)}+2 G_{(ij)(ik)} +G_{(ik)(ik)} \q ,
 \ea
where we denote with $A^2(a \wedge b)$ the area square of the parallelogram spanned by the vectors $a$ and $b$. Now we can expand the area squares and area metric variables into background (denoted by bold letters) and perturbations (denoted by small greek letters)
$
A^2={\bf A}^2+\alpha$ and $G={\bf G} + {}^A\!\mu^+
$, and we identify 
\ba
\alpha(e_i \wedge e_j)&\equiv& \alpha^+_{(2^i, 2^j)}   \q  \text{for}\, i<j \,, \nn\\
\alpha(e_i \wedge (e_j+e_k)) &\equiv& \alpha^+_{(2^i,2^j+2^k)} \q  \text{for}\,i\neq j,k \, \text{and}\,  j<k\,.
\ea
Equation (\ref{AMAtrafo1}) does then define a transformation between the fluctuations of the area squares $\alpha^+{(X,X'})$, defined in (\ref{trafo2}), and the fluctuations ${}^A\!\mu^+_{(ij)(km))}$ of the area metric, given by
\ba\label{AMAtrafo3}
\alpha^+_{(2^i,2^j)} &=& {}^A\!\mu^+_{(ij)(ij)}          \q \text{with}\,\,\, i<j \nn\\
\alpha^+_{(2^i,2^j+2^k)}&=& {}^A\!\mu^+_{O(ij) O(ij)}+ {}^A\!\mu^+_{O(ik) O(ik)} +  2\,\,{}^A\!\mu^+_{(ij) (ik)}      \q \text{with}\,\,i\neq j,k \, \text{and}\,  j<k \q . 
\ea
Here we indicate with $O(ij)$ the ordered index pair. We therefore  work with the following set of independent area metric fluctuations
\ba\label{MuPlusBasis}
&&\{  {}^A\!\mu^+_{(01)(01)}, {}^A\!\mu^+_{(02)(02)},{}^A\!\mu^+_{(03)(03)}, {}^A\!\mu^+_{(12)(12)},{}^A\!\mu^+_{(13)(13)},{}^A\!\mu^+_{(23)(23)},\nn\\
&&\,\, {}^A\!\mu^+_{(01)(02)}, {}^A\!\mu^+_{(02)(03)}, {}^A\!\mu^+_{(01)(03)}, 
 {}^A\!\mu^+_{(10)(12)}, {}^A\!\mu^+_{(10)(13)}, {}^A\!\mu^+_{(12)(13)}, \nn\\
&&\,\, {}^A\!\mu^+_{(20)(21)}, {}^A\!\mu^+_{(20)(23)}, {}^A\!\mu^+_{(21)(23)}, 
 {}^A\!\mu^+_{(30)(31)}, {}^A\!\mu^+_{(30)(32)}, {}^A\!\mu^+_{(31)(32)}
\}     \q .
\ea
For later use we will introduce the abstract index $B$ which can take values $B=(01)(01),(02)(02),\ldots, (31)(32)$.
We define the same transformation as in (\ref{AMAtrafo3}) for the Minus sector by replacing the $+$-super-index with the $-$-super-index everywhere.  
In (condensed) matrix notation we have
\ba
\begin{pmatrix}\alpha^+ \\
\alpha^-
\end{pmatrix}
 \,=\, {}^A\!{\mathbb M} 
 \cdot 
 \begin{pmatrix}
{}^A\!\mu^+ \\
  {}^A\!\mu^-
  \end{pmatrix} 
  \,=\, 
   \begin{pmatrix} {}^A\! M & 0 \\
   0 & {}^A\! M
   \end{pmatrix} 
   \cdot 
 \begin{pmatrix}
 {}^A\!\mu^+ \\
 {}^A\!\mu^-
  \end{pmatrix} 
\ea
where ${}^A\! M$ is a $18 \times 18$ matrix (which does not depend on the momenta $k$), see Appendix \ref{AppB}. Note that  ${}^A\! M$ is not a unitary matrix.

The Fourier transformed lattice Hessian  can be expressed in the $({}^A\mu^+, {}^A\mu^-)$ basis as 
\ba
H^{A\mu}(k')\,=\,       
{}^A\! {\mathbb M}^T \cdot  {\mathbb P}^T\!(-k')  \cdot H(k') \cdot {\mathbb P}(k')  \cdot {}^A\!{\mathbb M}    \q ,
\ea
 where ${\mathbb A}^T$ denotes the transpose of the matrix ${\mathbb A}$.

Expanding the transformed Hessian in powers of $\Lambda$, we see that the Hessian has different scaling behavior in the Plus- and Minus-sector:
\begin{itemize}
\item
The zeroth order coefficient of the Hessian decouples the Plus and Minus sector. There are 10 null vectors in the Plus-Plus block, whereas all eigenvalues of the Minus-Minus block are non-vanishing. We will see below that the null space of the Plus-Plus block is spanned by the length square fluctuations projected to the Plus sector. 
\item
More generally, the diagonal blocks, that is the Plus-Plus block and the Minus-Minus block contain only even powers in $\Lambda$ (and therefore in $k$) and the non-diagonal blocks only odd powers in $\Lambda$ (and therefore in $k$). 
\end{itemize}

\subsection{Identification of Length metric degrees of freedom}\label{Sec:Length}

We will proceed by defining yet another transformation, which isolates the projections to the Plus and Minus sector of the length square fluctuations. We will again see below, that the Plus part of the length square fluctuations is dynamically distinguished. 

Consider a geometry for the lattice defined by the length squares of its edges ${L}^2_e$. A perturbation of the length square of a given edge to $L^2_e={\bf L}^2_e+\delta$  will, to lowest order, induce a  perturbation  $A^2_t \approx {\bf A}^2_t+ (\ell_e)_t \, \delta$ for the area squares where
\ba
(\ell_e)_t  &=& \frac{ \partial A^2_{ t}}{\partial L^2_{e} }_{|L^2={\bf L}^2} \q ,
\ea
defines the components of a vector describing the area square fluctuations induced by a change of the length square of $e$.  
Here we can write the label for the triangles as $t=(\nu,\tilde t)$, where $\nu$ denotes a vertex and $\tilde t$ labels the 36 triangles which are associated to this  vertex and  are in the boundary of the hypercube associated to this vertex. Likewise we can write $e=(\nu,\tilde e)$, where $\tilde e$ labels the 14 edges associated to $\nu$  and are in the boundary of this hypercube associated to this vertex. These 14 edges are given by the 4 edges parallel to the principal axis vectors $e_i$ (e.g. for the vertex $\nu=\{0\}$ the edges $\{0,2^i\}$), 6 edges parallel to the face diagonals $e_i+e_j$ (i.e. the edges $\{0,2^i+2^j\}$), and 4 edges parallel to the body diagonals $e_i+e_j+e_k$  (i.e. the edges $\{0,2^i+2^j+2^k\}$), where $i,j,k=0,\ldots,3$ and $i,j,k$ are mutually different.  However, in our lattice background geometry, a given body diagonal is the hypothenuse for all the triangles (in the boundary of some hypercube), it is contained in. Thus $\partial A^2_{(\nu,\tilde t)}/\partial L^2_{(\nu',\tilde e)}$ is vanishing, if $\tilde e$ denotes a body diagonal.  The lengths fluctuations $\ell_{(\nu,e)}$ therefore lead to only 10 independent combinations of area square fluctuations.  In the following we will restrict the index $\tilde e$ to label the principal axes and the face diagonals. Fortunately, these 10 variables are necessary and sufficient to construct the length metric fluctuations. 

We can express the (Fourier transformed) $(\ell_{\tilde e}(k'))$ in the $({}^A\!\mu^+,{}^A\!\mu^-)$ basis and obtain
\ba
\ell^{A\mu}_{\tilde e}(k') &=&  ({}^A\! {\mathbb M})^{-1} \cdot  ({\mathbb P}(k'))^{-1} \cdot  \ell_{\tilde e}(k')  \q .
\ea
It turns out that the components of $\ell^{A\mu}_{\tilde e}(k')$ for a given $\tilde e \equiv \{0,X\}$ have a simple momentum dependence: For the Plus sector this is given by the factor $(1+\omega^{-1}_X)$, which includes a $\Lambda^0$ term. For the Minus sector this is given by the factor $(1-\omega^{-1}_X)$, which is of order $\Lambda^1$.  (Note that this feature does depend on the particular choice of $\omega$ factors made in (\ref{trafo2}).)

That is we can write 
\ba
\ell^{A\mu}_{\{0,X\}}(k')  &=&  (\omega^{-1}_X+1)(k')\, \,\ell^{A\mu+}_{\{0,X\}} +  (\omega^{-1}_X-1)(k') \,\, \ell^{A\mu-}_{\{0,X\}}
\ea
where $X=2^i$ or $X=2^i+2^j$, and the components $\ell^{A\mu+}_{\{0,X\}}$ and $\ell^{A\mu-}_{\{0,X\}}$ do not depend on $k'$. 

The 10 (independent) length vectors  $\ell^{A\mu}_{\tilde e}$ lead in this way to 10 independent vectors $\ell^{A\mu+}_{\{0,X\}}$ in the Plus sector. But the 10 vectors $\ell^{A\mu-}_{\{0,X\}}$ in the Minus sector turn out to span only a 9-dimensional space\footnote{
The $\ell^{A\mu-}_{\{0,X\}}$ turn out to satisfy the equation $\sum_i\ell^{A\mu-}_{\{0,2^i\}}+2 \sum_{i<j}\ell^{A\mu-}_{\{0,2^i+2^j\}}=0$.
}. 
We can build orthogonal\footnote{with respect to the canonical inner product defined by the basis (\ref{MuPlusBasis}) } projectors ${}^L\!\Pi^+$ and ${}^L\!\Pi^-$, which project onto the space spanned by the sets $\{\ell^{A\mu+}_{\{0,X\}}\}$ and $\{ \ell^{A\mu-}_{\{0,X\}} \}$, respectively. See Appendix \ref{AppA} for a straightforward procedure, which also allows for over-complete sets of vectors. We can then use ${\mathbb I}^+-{}^{L}\!\Pi^+$ and ${\mathbb I}^--{}^{L}\!\Pi^-$ to identify the degrees of freedom which cannot be generated by the length perturbation in the way described above. Here we denote with ${\mathbb I}^+$ and ${\mathbb I}^-$ the projection to the Plus and Minus sector respectively.

The columns of ${\mathbb I}^\pm-{}^{L}\!\Pi^\pm$ define an over-complete set of vectors, orthogonal to the set $\{\ell^{A\mu\pm}_{\{0,X\}}\}$. We could extract a set of 8 (for the Plus sector) or 9 (for the Minus sector) independent vectors defining the same span as the sets of these columns. This will however introduce arbitrary choices, and it is more convenient to express the Hessian in the over-complete basis. In fact, this just amounts to apply the projectors to the Hessian.  This will also save us from introducing yet another index, as the columns of the projector are labelled by the same kind of index $B$ as introduced below (\ref{MuPlusBasis}). We therefore introduce the vectors $\chi^\pm_B$ defined by. 
\ba
(\chi^\pm_B)_{\mathbb{B}'}=({\mathbb I}^\pm-{}^{L}\!\Pi^\pm)_{\mathbb{B}' B} \q ,
\ea
where $\mathbb{B}'$ runs through a Plus sector copy and a Minus sector copy, that is it can take on 36 values; but $B$ rans through only one sector, that is it can take on 18 values.
These $\chi$-vectors are closely related to the hyper-cubical constraints, which appeared for the standard triangulation of the hypercubic lattice in \cite{ARE1} (and where also named $\chi$ there). 

We now have an over-complete basis formed by the  $\ell^{A\mu+}_{\{0,X\}},\ell^{A\mu-}_{\{0,X\}},\chi^+_B$ and $\chi^-_B$ vectors, and could evaluate the Hessian in this basis. However, we will rather need the Hessian contracted with the (momentum dependent) length vectors $\ell^{A\mu}$. We will even obtain simpler expressions, if we transform the length vectors to length metric vectors, that is consider perturbations of the area squares which are induced by (a lattice discretization) of the length metric. 

To this end we note that the transformation from length square to length metric variables proceeds very similar as from area square to area metric variables, discussed in Section \ref{Sec:separation}, and leads to a $10\times 10$ transformation matrix $({}^L\! M)_{\{0,X\} I}$ ( see Appendix \ref{AppB}). Here $X=2^i$ or $X=2^i+2^j$ for $i,j=0,\ldots,3$ and $i<j$ and $I $ can take values 
\ba\label{IndexI}
I=00,11,22,33,01,02,03,12,13,23 \q ,
\ea
 so that it labels the independent components of the length metric.
 
 We can thus introduce the vectors
 \ba
\eta_I (k') &=&( \ell^{A\mu}(k') \cdot {}^L\! M   )_I \q\;\,\,=\, \sum_{X} \ell^{A\mu}_{0X}(k') \,  ({}^L\! M)_{\{0,X\} I} \q ,\nn\\
\zeta^-_I &=& \lambda^{-2} ( \ell^{A\mu-} \cdot {}^L\! M   )_I\,\,\,\,\,\,=\, \lambda^{-2}\sum_{X} \ell^{A\mu-}_{0X} \,  ({}^L\! M)_{\{0,X\} I} \q ,
 \ea
and use these instead of  the $\ell^{A\mu+}_{\{0,X\}}$ and $\ell^{A\mu-}_{\{0,X\}}$ vectors. (The $\zeta^-_I$ vectors are closely related to the $\zeta$-vectors introduced in \cite{ARE1}.)

We introduced the length metric vectors $\eta_I (k')$ only now, as they do {\it not} have the same kind of homogeneous momentum dependence as the length vectors $\ell^{A\mu}_{\{0,X\}}$. This homogeneous momentum dependence of the $\ell^{A\mu}_{\{0,X\}}$ allowed us to construct the (momentum independent) projectors ${}^L\!\Pi^+$ and ${}^L\!\Pi^-$, and thus  (momentum independent) $\zeta^-$-vectors. 

Note that the length metric vectors $\eta_I (k')$ are build from a Plus sector component and a Minus sector component. But the norm of the Plus sector component scales with $\Lambda^0$, whereas the norm of the Minus sector component scales with $\Lambda^1$. Thus the Minus sector components can be considered to be sub-leading. In the following we will interpret the length metric degrees of freedom as part of the Plus sector. 
%Note that both the zeroth order of the Hessian contracted with the $\eta$-vectors as well as the zeroth order of the Hessian contracted with vectors $\ell^{A\mu+}\cdot {}^L\! M$ vanishes.

We can derive the projectors ${}^L\!\Pi^+$ and thus ${\mathbb I}^+-{}^L\!\Pi^+$ also from the continuum expression for an area metric $G_{(ij)(kl)}$ induced by a length metric $g_{ik}$.  This expression is given by
\ba\label{CARL1}
G_{(ij)(kl)}&=& g_{ik}g_{jl}-g_{il}g_{jk}      \q .
\ea
If we expand both metrics into background and perturbations, $G={\mathbf G}+ {}^A\!\mu^+$ and $g={\mathbf g}+ {}^L\!\mu^+$, we obtain an expression for the area metric fluctuations induced by length metric fluctuations
\ba\label{CARL2}
{}^A\!\mu^+_{(ij)(kl)}&=& \lambda^2 ( \delta_{ik} {}^L\!\mu^+_{jl}+ \delta_{jl}{}^L\!\mu^+_{ik}-\delta_{il} {}^L\!\mu^+_{jk} -\delta_{jk}{}^L\!\mu^+_{il})
\ea
Restricting the indices $(ij)(kl)$ to be of type $B$ (see (\ref{MuPlusBasis})) and taking the derivative of the RHS of (\ref{CARL2}) with respect to ${}^L\!\mu^+_I$, where $I$ runs through the values (\ref{IndexI}), we obtain 10 vectors, that span the space of Area metric fluctuations induced by Length metric fluctuations. These 10 vectors agree with 
$\sum_{X=2^k,2^k+l}\ell^{A\mu+}_{\{0,X\}}( {}^L\! M)_{\{0,X\}I}$, that is they span the same space as the $\ell^{A\mu+}_{\{0,X\}}$. The projector onto the space spanned by these 10 vectors will therefore agree with the restriction of ${}^L\!\Pi^+$ to the Plus sector.  We will write this restriction as  $(P^+)^\dagger \cdot {}^L\!\Pi^+ \cdot P^+$, where $P^+$ is a $36\times 18$ matrix, whose diagonal entries are 1, and with all other entries vanishing. The projector $ \chi^+_+\,=\,(P^+)^\dagger \cdot ({\mathbb I}^+-{}^L\!\Pi^+) \cdot P^+$ is given by 
\ba
\chi^+_+=\frac{1}{6}
{\tiny
\left(
\begin{array}{cccccccccccccccccc}
 2 & -1 & -1 & -1 & -1 & 2 & 0 & 0 & 0 & 0 & 0 & 0 & 0 & 0 & 0 & 0 & 0 & 0 \\
 -1 & 2 & -1 & -1 & 2 & -1 & 0 & 0 & 0 & 0 & 0 & 0 & 0 & 0 & 0 & 0 & 0 & 0 \\
 -1 & -1 & 2 & 2 & -1 & -1 & 0 & 0 & 0 & 0 & 0 & 0 & 0 & 0 & 0 & 0 & 0 & 0 \\
 -1 & -1 & 2 & 2 & -1 & -1 & 0 & 0 & 0 & 0 & 0 & 0 & 0 & 0 & 0 & 0 & 0 & 0 \\
 -1 & 2 & -1 & -1 & 2 & -1 & 0 & 0 & 0 & 0 & 0 & 0 & 0 & 0 & 0 & 0 & 0 & 0 \\
 2 & -1 & -1 & -1 & -1 & 2 & 0 & 0 & 0 & 0 & 0 & 0 & 0 & 0 & 0 & 0 & 0 & 0 \\
 0 & 0 & 0 & 0 & 0 & 0 & 3 & 0 & 0 & 0 & 0 & 0 & 0 & 0 & 0 & 0 & 0 & -3 \\
 0 & 0 & 0 & 0 & 0 & 0 & 0 & 3 & 0 & 0 & 0 & 0 & 0 & 0 & -3 & 0 & 0 & 0 \\
 0 & 0 & 0 & 0 & 0 & 0 & 0 & 0 & 3 & 0 & 0 & -3 & 0 & 0 & 0 & 0 & 0 & 0 \\
 0 & 0 & 0 & 0 & 0 & 0 & 0 & 0 & 0 & 3 & 0 & 0 & 0 & 0 & 0 & 0 & -3 & 0 \\
 0 & 0 & 0 & 0 & 0 & 0 & 0 & 0 & 0 & 0 & 3 & 0 & 0 & -3 & 0 & 0 & 0 & 0 \\
 0 & 0 & 0 & 0 & 0 & 0 & 0 & 0 & -3 & 0 & 0 & 3 & 0 & 0 & 0 & 0 & 0 & 0 \\
 0 & 0 & 0 & 0 & 0 & 0 & 0 & 0 & 0 & 0 & 0 & 0 & 3 & 0 & 0 & -3 & 0 & 0 \\
 0 & 0 & 0 & 0 & 0 & 0 & 0 & 0 & 0 & 0 & -3 & 0 & 0 & 3 & 0 & 0 & 0 & 0 \\
 0 & 0 & 0 & 0 & 0 & 0 & 0 & -3 & 0 & 0 & 0 & 0 & 0 & 0 & 3 & 0 & 0 & 0 \\
 0 & 0 & 0 & 0 & 0 & 0 & 0 & 0 & 0 & 0 & 0 & 0 & -3 & 0 & 0 & 3 & 0 & 0 \\
 0 & 0 & 0 & 0 & 0 & 0 & 0 & 0 & 0 & -3 & 0 & 0 & 0 & 0 & 0 & 0 & 3 & 0 \\
 0 & 0 & 0 & 0 & 0 & 0 & -3 & 0 & 0 & 0 & 0 & 0 & 0 & 0 & 0 & 0 & 0 & 3 \\
\end{array}
\right)  \q .
}
\ea
From this explicit form of $\chi^+_+$  one can see that it projects out from an area metric fluctuation ${}^A\!\mu_{(ij)(kl)}$ all possible traces. %That is %$\sum_{ik} ( \chi^+_+ \cdot {}^A\!\mu)_{(ij)(kl)} %\delta^{ik}=0$ and $\sum_{jl} ( \chi^+_+ \cdot %{}^A\!\mu)_{(ij)(kl)} \delta^{jl}=0$.

Note that the projector ${}^L\!\Pi^-$ does not agree with ${}^L\!\Pi^+$, as indeed they project onto spaces of dimension $9$ and $10$ respectively.  The matrix expression for ${}^L\!\Pi^-$ is considerably more involved than the one for ${}^L\!\Pi^+$, see Appendix \ref{AppC}, and depends on choices of signs in the definition of the $\alpha^-$ variables in (\ref{trafo2}).

\section{Accessing the effective dynamics for the Length metric degrees of freedom}\label{AccEff}

In this section we will extract the two leading order terms for the effective dynamics of the length metric. The leading order term will be given by the Hessian block for the length metric fluctuations itself. This Hessian block provides a discretization of the linearized Einstein-Hilbert action, see Section \ref{LengthBlock}.  Corrections for this dynamics will however arise, when we integrate out the remaining degrees of freedom, see Section \ref{Sec:EffH}. We will  show in Section \ref{Sec:Scaling} that the Plus and Minus sector have different scaling behavior in the Area Regge action. To determine the lowest order correction in the lattice constant, we only need to consider the Plus sector. We will then compute this correction in Section \ref{Sec:Low} and state a number of its properties. We will provide a geometrical interpretation of the correction terms in Section \ref{Sec:Weyl}, and identify it as a quadratic contraction in the linearized Weyl curvature tensor.

 \subsection{The Hessian block for the length metric fluctuations}\label{LengthBlock}

The previous subsection yielded the definition of an over-complete basis of vectors $\{ \eta_I, \chi^+_B, \zeta^-_I, \chi^-_B\}$. Here we have $I$ labelling the independent components of the length metric, see (\ref{IndexI}) and $B$ labelling the independent components of the area metric, see (\ref{MuPlusBasis}). The $\eta_I$ vectors describe fluctuations in the area squares, which are induced by a length metric perturbation. All other vectors describe degrees of freedom for geometrical excitations more general than the ones defined by a length metric. We will see that we can interpret the $\eta$ and $\chi^+$ vectors to encode the area metric perturbations (of the 18 independent components available after integrating out the triangles in the bulks of the hypercubes). $\zeta^-_I, \chi^-_B$ seem to describe more microscopic degrees of freedom. Indeed, they arise from the differences of area square perturbations associated to triangles in the same plane, and we will see that integrating out these degrees of freedom does not affect the effective Lagrangian for the length metric degrees of freedom up to the $\Lambda^4$-th order.

We will first consider the Hessian contracted with the length metric vectors $\{\eta_I\}$ on both sides. We denote by $\eta$ the $36\times 10$ matrix obtained by using the $\eta_I$ as its columns. The contraction of the Hessian with the $\eta_I$  will lead to a discretization of the linearized Einstein-Hilbert action. To write this in a compact form we introduce spin zero mode and spin two mode projections 
${}^0\!\Pi_{\rm L},{}^2\!\Pi_{\rm L}$ on the lattice \cite{BahrDittrichHe}, see Appendix \ref{AppE}. We then have 
\ba\label{LEHL}
H_{\eta\eta}(k)\,:=\,\eta^T(-k)  \cdot H^{A\mu} \cdot \eta(k)&=&  -\frac{1}{4}\Delta_{\rm L} ({}^2\!\Pi_{\rm L}-2\,{}^0\!\Pi_{\rm L})
\ea
with 
\ba
\Delta_{\rm L}=\frac{1}{\lambda^2}\sum_{i=0}^3 \left(2-\exp(\imath \Lambda k_i^2)-\exp(-\imath \Lambda k_i^2)\right)\,\,=\, \, \frac{\Lambda^2}{\lambda^2}\sum_{i=0}^3 k_i^2 + \frac{1}{\lambda^2} {\cal O}(\Lambda^3)
\ea
 the lattice Laplacian. Note that, after setting $\Lambda=\lambda$, the contribution of (\ref{LEHL}) to the action is of zeroth order in the lattice constant $\lambda$. We will see, that this will turn out to be the leading order in the effective action for the length metric, in the continuum limit $\lambda\rightarrow 0$.

This result is on the one hand not surprising: contracting the Area Regge Hessian with vectors $\partial A^2_t/\partial L_e^2$ from both sides one indeed obtains the Length Regge Hessian. This leads also to the recovery of the linearized Einstein-Hilbert action %\footnote{In the case of the Length Regge action for the standard triangulation of the hypercubic lattice, one has to integrate out the lengths of the body diagonals to arrive at the linearized Einstein-Hilbert action \cite{RocekWilliams}.} 
 for the standard triangulation of the hypercubic lattice \cite{ARE1}. 
But different from the standard triangulation, we have here (before contracting with the Length metric vectors) integrated out the area squares of the triangles which are bulk in any hypercube: this involved 64 area variables for each hypercube and therefore 64 variables per lattice vertex.  We see that we can nevertheless extract the linearized Einstein-Hilbert action after integrating out all these area variables, which are not constrained in any way to result from length fluctuations.

In fact, if one does expand the Length Regge action on the standard triangulation of the hypercubic lattice, one needs, in order to obtain the linearized Einstein-Hilbert action, to integrate out the length squares associated to the body diagonals \cite{RocekWilliams}. The reason is that these turn out to be massive degrees of freedom, and are in fact `spurious', that is not needed for a matching between the length and length metric fluctuation. %This integrating out procedure imposes that certain deficit angles associated to bulk triangles in the hypercube vanish \cite{WilliamsRoceck}. 

Here we have already integrated out the area variables associated to the bulk triangles in the hypercube. The length squares of the body diagonal are only affected by variations of the area squares associated to the bulk triangles. This integrating out procedure has therefore removed the body diagonal lengths from our set of variables already.

\subsection{Computing effective Hessians}\label{Sec:EffH}

We have seen that the linearized  Area Regge action, after integrating out the bulk degrees of freedom for all hypercubes, reduces for the length metric subset of variables to the linearized Einstein-Hilbert action. But apart from the 10 length metric variables per vertex, we have additionally 26 further variables, described by the subset of (over-complete) basis vectors $\{\chi^+_B, \zeta^-_I, \chi^-_B\}$.  We can integrate out these degrees of freedom, to obtain an effective action for the length metric variables only. The scaling properties of the Hessian with respect to $\Lambda$ in its various blocks will tell us  whether the linearized Einstein-Hilbert action for the length metric block will be also the dominant one for the effective action. 

Consider a free theory with variables $y\equiv \{y_a\}_{a=1}^{N_y}$ and $z\equiv \{z_b\}_{b=1}^{N_z}$ and an action
\ba
S =\frac{1}{2}\left(y^\dagger \cdot H_{yy} \cdot y  +  y^\dagger \cdot H_{yz} \cdot z + z^\dagger \cdot H_{zy} \cdot y + z^\dagger \cdot H_{zz} \cdot z\right) \q .
\ea
Integrating out the $z$-degrees of freedom amounts to solving the equations of motion for the $z$-variables and inserting the solutions back into the action. If $H_{zz}$ is invertible, the resulting effective action for the $y$-variables is given by
\ba\label{I1}
S'=\frac{1}{2}\left(\, y^\dagger \cdot(H_{yy}- H_{yz} \cdot H_{zz}^{-1} \cdot H_{zy}) \cdot y \, \right) \q .
\ea
$H_{zz}$ may not be invertible, but rather have null vectors $n_A$, labelled by $A=1,\ldots$. If this is due to gauge symmetries, the vectors $n_A$ will be also null with respect to $H_{yz}$. We can then replace $H_{zz}^{-1}$ in (\ref{I1}) by $(H_{zz} + {\cal G})^{-1}$ where 
\ba\label{GF01}
 {\cal G} \,=\, \sum_{A}  \beta_A \,  (n_A)(n_A)^\dagger       \q .
\ea
Note that using an over-complete basis can be seen as introducing gauge symmetries. In other words, if we have an over-complete basis $\{Z_b\}_{b=1}^{N_z}$ defining the $z$-variables, we have coefficients $V_B^b$, which satisfy $ \sum_b Z_b V_B^b =0$. The $\{V_B\}_B$ are then null vectors for $H_{zz}$ and $H_{yz}$, and hence can be understood to parametrize the gauge directions for the $z$-variables.
This allows us to work with the over-compete basis vectors $\{\chi^+_B, \zeta^-_I, \chi^-_B\}$.  The various diagonal Hessian  blocks can be computed as
\ba\label{Bl1}
H_{\chi^\pm \chi^\pm}(k)\,=\, (\chi^\pm)^T \cdot  H^{A\mu}(k) \cdot \chi^\pm\q ,\q\q H_{\zeta^-\zeta^-}(k)\,=\, \zeta^T \cdot  H^{A\mu}(k) \cdot \zeta \q ,
\ea
whereas the non-diagonal blocks are given by
\ba\label{Bl2}
H_{\eta \chi^\pm}(k)\,=\, \eta(-k) \cdot  H^{A\mu}(k) \cdot \chi^\pm \q , \q\q H_{\chi^\pm\eta }(k)\,=\, \chi^\pm \cdot  H^{A\mu}(k) \cdot \eta(k)
\ea
and so on. Note that, e.g. $(H_{\eta \chi^\pm}(k))^\dagger=H_{\chi^\pm\eta }(k)$.

 \subsection{Scaling behavior for the Hessian blocks}\label{Sec:Scaling}

Here we will state the scaling behavior in $\lambda$ and  $\Lambda$ of the various Hessian blocks defined in (\ref{Bl1},\ref{Bl2}). The scaling behavior in $\lambda$ results from a global factor of $\lambda^{-6}$ in the Hessian $H^{A\mu}$ (which can be determined from dimensional analysis), and the fact that the $\eta_I$ include a factor of $\lambda^2$ from the Area square -- Length square derivative. All other basis vectors scale with $\lambda^0$. 

The scaling behavior in $\Lambda$ can be determined by expanding the matrix elements of the Hessian blocks in $\Lambda$; more precisely we expand the $\omega$ and $\omega^{-1}$ factors in terms of $\Lambda$. For each block we will only consider the lowest order in $\Lambda$. Note that  all basis vectors have a norm of order $\Lambda^0$.

One finds the following scaling for the Hessian blocks:
\ba\label{Scaling1}
H^{A\mu}\,\sim \q \frac{1}{\lambda^6}\,\,\times\,\,
\begin{tabular}{|c |c|c|c|c|}
\hline
&$\eta$ & $\chi^+$ &$ \zeta^-$ & $\chi^-$ \\   \hline
$\eta$ & $\;\; \lambda^4{\cal O}(\Lambda^2)\;\;$ & $ \;\;\lambda^2  {\cal O}(\Lambda^2)\;\; $ &  $ \;\;\lambda^2  {\cal O}(\Lambda^3)\;\; $ & $\;\; \lambda^2  {\cal O}(\Lambda^3) \;\; $ \\ \hline
$\chi^+$ & $ \lambda^2  {\cal O}(\Lambda^2)$ & $\lambda^0{\cal O}(\Lambda^0)$ & $\lambda^0{\cal O}(\Lambda^1)$ & $\lambda^0{\cal O}(\Lambda^1)$\\ \hline
$\zeta^-$ & $\lambda^2  {\cal O}(\Lambda^3)$& $\lambda^0{\cal O}(\Lambda^1)$ & $ \lambda^0{\cal O}(\Lambda^0)$ & $\lambda^0{\cal O}(\Lambda^0)$\\\hline
$\chi^-$ & $\lambda^2  {\cal O}(\Lambda^3)$ & $\lambda^0{\cal O}(\Lambda^1)$ & $\lambda^0{\cal O}(\Lambda^0)$ & $ \lambda^0{\cal O}(\Lambda^0)$ \\ \hline
\end{tabular}              \q .
\ea
Note that in particular the $\Lambda^0$-part of the Hessian, has, apart from the null vectors induced by the over-completeness of the basis, only the Length-metric degrees of freedom as null modes. We can thus characterize these degrees of freedom as massless, whereas all other degrees of freedom come with a lattice constant dependent mass. (The 10 metric degrees of freedom do include 4 null vectors for the $H_{\eta \eta}$ block, that describe the diffeomorphism modes, and can be extended to the full Hessian. These 4 diffeomorphism modes induce 4 constraints, leaving the usual two spin-2 degrees of freedom associated to the graviton.)

We also see that the Plus sector and the Minus sector are dynamically distinguished: firstly, the $\Lambda^0$-part of the Hessian decouples the Plus and Minus sector. Secondly, the non-diagonal blocks between the length metric variables and the remaining Plus sector variables are of order ${\cal O}(\Lambda^2)$, whereas between the length metric variables and the minus sector variables it is ${\cal O}(\Lambda^3)$.

With this scaling behavior of the Hessian blocks, we can see that integrating out the Minus sector variables will lead to a correction $K^-$ for the Hessian for the Plus sector, whose lowest order is given by
\ba\label{Scaling2}
K^-\sim \q \frac{1}{\lambda^6}\,\,\times \,\,
\begin{tabular}{|c |c|c|} \hline
&$\eta$          & $\chi^+$ \\\hline
$\eta$ & $\lambda^4 {\cal O}(\Lambda^6)$  & $\lambda^2 {\cal O}(\Lambda^4)  $ \\\hline
$\chi^+$ & $\lambda^2 {\cal O}(\Lambda^4)$  &$ \lambda^0 {\cal O}(\Lambda^2) $ \\\hline
\end{tabular}         \q .
\ea
Note that all blocks in $K^{-}$ are sub-leading as compared to the corresponding blocks in the Hessian $H^{A\mu}$.

Thus, if we integrate out in a second step the $\chi^+$ variables, the lowest order corrections come only from the Plus-Plus block of $H^{A\mu}$,  whereas the correction $K^{-}$ does  contribute to higher order. This lowest order correction, coming from the $\chi^+$ variables, will be of order $\lambda^{-2}{\cal O}(\Lambda^4)$.

Here we will compute only the lowest order modification for the graviton dynamics. In the following we therefore restrict to the Plus-Plus block of the Hessian. This Plus-Plus  block of the Hessian can be interpreted to define a linearized action for the area metric fluctuations.

\subsection{Lowest order correction in the effective action for the length metric fluctuations}\label{Sec:Low}

Here we consider the lowest order correction for the effective Hessian of the length metric variables, which results from integrating out the $\chi^+$ variables:
\ba\label{LO1}
K^{\chi^+}&:=& - H_{\eta \chi^+} \cdot (H_{\chi^+ \chi^+}+{\cal G}_{\chi^+ \chi^+})^{-1}  \cdot H_{\chi^+ \eta}  \q ,
\ea
where we denote with ${\cal G}_{\chi^+ \chi^+}$ the gauge fixing term of the form (\ref{GF01}), which we need to add due to the over-completeness of the $\{\chi^+_B\}_B$-basis. 

To compute the $\Lambda^4$ order of $K^{\chi^+}$, we need to consider only the lowest $\Lambda$-orders in $H_{\eta \chi^+}, H_{\eta \chi^+}$ and the inverse of $(H_{\chi^+ \chi^+}+{\cal G}_{\chi^+ \chi^+})$. That is the $\Lambda^2$ terms for $H_{\eta \chi^+}, H_{\eta \chi^+}$ and the $\Lambda^0$ term for $(H_{\chi^+ \chi^+}+{\cal G}_{\chi^+ \chi^+})$.

It turns out that we can obtain an astonishingly simple expression for the correction, if we slightly re-organize the various contractions in (\ref{LO1}). To this end, we remember that  the $36 \times 18$ matrix $\chi^+$ formed by taking the $\chi^+_B$-vectors as its columns, define the first 18 columns of the projector $({\mathbb I}^+-{}^{L}\!\Pi^+)$, see Section \ref{Sec:Length}.  We defined $\chi^+_+:=P^+\chi^+$ where $P^+$  acts as restriction map to the Plus sector. $\chi^+_+$ is a symmetric, orthogonal projector. 

We can therefore write the correction $K^{\chi^+}$ also as  
\ba
K^{\chi^+}(k)&:=& - \,\,\left[\eta(-k) \cdot H^{A\mu}(k) \cdot {P}^+\right]\, \cdot \,\left[  \chi^+_+ \cdot (H_{\chi^+ \chi^+}(k)+{\cal G}_{\chi^+ \chi^+})^{-1}  \cdot \chi^+_+ \right]\cdot \left[  ( P^+)^\dagger \cdot H^{A\mu}(k) \cdot \eta(k) )\right] \nn\\
&=:&\q\q\q\q\q\q\q-\q \;{\cal R}^\dagger \q \q\q\q \cdot\q \q \q\q {\cal M}\q\q\q \q \cdot \q\q\q \q {\cal R}
\ea

The reason for doing so, is that we can provide the elements of the matrices in square brackets in a more compact form as compared to the matrices in (\ref{LO1}). Additionally, by shifting the projectors $\chi_+^+$  from $H_{\eta\chi^+}$ and $H_{\chi^+\eta}$ onto  the central matrix in (\ref{LO1}), the resulting matrix  ${\cal M}$ does not depend any more on the gauge parameters appearing in (\ref{LO1}).  We remind the reader that the indices $B$ and $I$ take values
\ba\label{IndexV}
B=&&(01)(01),(02)(02),(03)(03),(12)(12),(13)(13),(23)(23), \nn\\
&&(01)(02), (02)(03), (01)(03), 
(10)(12), (10)(13),(12)(13), \nn\\
&&(20)(21), (20)(23), (21)(23), 
(30)(31), (30)(32), (31)(32) \, ; \nn\\
I=&&00,11,22,33,01,02,03,12,13,23\, ;
\ea
respectively.  The matrix ${\cal R}$ can then be computed to be
\ba
{\cal R}={\tiny \sqrt{2} \frac{\Lambda^2}{\lambda^4}
\left(
\begin{array}{cccccccccc}
 0 & 0 & \frac{k_3^2}{2} & \frac{k_2^2}{2} & 0 & 0 & 0 & 0 & 0 & -k_2 k_3 \\
 0 & \frac{k_3^2}{2} & 0 & \frac{k_1^2}{2} & 0 & 0 & 0 & 0 & -k_1 k_3 & 0 \\
 0 & \frac{k_2^2}{2} & \frac{k_1^2}{2} & 0 & 0 & 0 & 0 & -k_1 k_2 & 0 & 0 \\
 \frac{k_3^2}{2} & 0 & 0 & \frac{k_0^2}{2} & 0 & 0 & -k_0 k_3 & 0 & 0 & 0 \\
 \frac{k_2^2}{2} & 0 & \frac{k_0^2}{2} & 0 & 0 & -k_0 k_2 & 0 & 0 & 0 & 0 \\
 \frac{k_1^2}{2} & \frac{k_0^2}{2} & 0 & 0 & -k_0 k_1 & 0 & 0 & 0 & 0 & 0 \\
 0 & 0 & 0 & -k_1 k_2 & 0 & 0 & 0 & -k_3^2 & k_2 k_3 & k_1 k_3 \\
 0 & 0 & -k_1 k_3 & 0 & 0 & 0 & 0 & k_2 k_3 & -k_2^2 & k_1 k_2 \\
 0 & -k_2 k_3 & 0 & 0 & 0 & 0 & 0 & k_1 k_3 & k_1 k_2 & -k_1^2 \\
 0 & 0 & 0 & -k_0 k_2 & 0 & -k_3^2 & k_2 k_3 & 0 & 0 & k_0 k_3 \\
 0 & 0 & -k_0 k_3 & 0 & 0 & k_2 k_3 & -k_2^2 & 0 & 0 & k_0 k_2 \\
 -k_2 k_3 & 0 & 0 & 0 & 0 & k_0 k_3 & k_0 k_2 & 0 & 0 & -k_0^2 \\
 0 & 0 & 0 & -k_0 k_1 & -k_3^2 & 0 & k_1 k_3 & 0 & k_0 k_3 & 0 \\
 0 & -k_0 k_3 & 0 & 0 & k_1 k_3 & 0 & -k_1^2 & 0 & k_0 k_1 & 0 \\
 -k_1 k_3 & 0 & 0 & 0 & k_0 k_3 & 0 & k_0 k_1 & 0 & -k_0^2 & 0 \\
 0 & 0 & -k_0 k_1 & 0 & -k_2^2 & k_1 k_2 & 0 & k_0 k_2 & 0 & 0 \\
 0 & -k_0 k_2 & 0 & 0 & k_1 k_2 & -k_1^2 & 0 & k_0 k_1 & 0 & 0 \\
 -k_1 k_2 & 0 & 0 & 0 & k_0 k_2 & k_0 k_1 & 0 & -k_0^2 & 0 & 0 \\
\end{array}
\right) + {\cal O}(\Lambda^3) \q .
}
\ea
We can also give its matrix elements in terms of a simple combination of Levi-Civita symbols. 
To this end we define for $B=(ij)(km),$  $B_1=i,B_2=j,B_3=k,B_4=m$. Likewise, for $I=ij$ we define $I_1=i$ and $I_2=j$. With these preparations we can write the matrix elements of $\left[  ( R^+ )^\dagger\cdot H^{A\mu}(k) \cdot \eta(k) \right]$ as  
\ba\label{3.14}
%\left[  ( R^+ )^\dagger\cdot H^{A\mu}(k) \cdot \eta(k) \right]_{BI}
{\cal R}_{BI}
= \frac{{\cal S}_{\rm AM}(B){\cal S}_{LM}(I)}{2\sqrt{2}} \frac{\Lambda^2}{\lambda^4}\sum_{m,n=0}^3 (\varepsilon_{B_1B_2I_1m} \varepsilon_{B_3 B_4I_2n} + \varepsilon_{B_1B_2I_2 m} \varepsilon_{B_3 B_4I_1 n}) k_mk_n  +{\cal O}(\Lambda^3) \, ,\q\q
\ea
where $\varepsilon_{ijmn}$ is the Levi-Civita symbol. ${\cal S}_{\rm AM}(B)$ and ${\cal S}_{\rm LM}(I)$ are factors of 2, that arise due to our summation convention for the area and length metric indices
\ba
{\cal S}_{\rm AM}(B)=
 \begin{cases}
 1 \q \text{for}\, (B_1B_2)=(B_3B_4) \\
 2 \q \text{for}\, (B_1B_2)\neq(B_3B_4)
 \end{cases}
 \q, \q\q
  {\cal S}_{\rm LM}(I)=
 \begin{cases}
 1 \q \text{for}\, I_1=I_2 \\
 2 \q \text{for}\, I_1\neq I_2
 \end{cases} \q .
\ea
We have defined $B$ to run only through the values listed in (\ref{IndexV}), which do not include  (01)(23),(02)(31),(03)(12). We can extend the expression (\ref{3.14}), allowing for $B$ to take on also these values. In this case we have also to extend the index range for ${\cal M}$ accordingly, and require all entries of ${\cal M}$ associated with these indices to be vanishing.

With this extension the matrix ${\cal M}$ is, modulo terms of order $\Lambda^2$ and higher, given by \ba
%&&
%\left[  \chi^+_+ \cdot (H_{\chi^+ \chi^+}(k)+{\cal G}_{\chi^+ \chi^+})^{-1}  \cdot \chi^+_+ \right] \,=\,\nn\\
%&&
{\cal M}\simeq {\tiny
-\tfrac{\lambda^6}{3 \cdot 4^2}
\left(
\begin{array}{ccccccccccccccccccccc}
 \frac{4}{9} & -\frac{2}{9} & -\frac{2}{9} & -\frac{2}{9} & -\frac{2}{9} & \frac{4}{9} & 0 & 0 & 0 & 0 & 0 & 0 & 0 & 0 & 0 & 0 & 0 & 0 & 0 & 0 & 0 \\
 -\frac{2}{9} & \frac{4}{9} & -\frac{2}{9} & -\frac{2}{9} & \frac{4}{9} & -\frac{2}{9} & 0 & 0 & 0 & 0 & 0 & 0 & 0 & 0 & 0 & 0 & 0 & 0 & 0 & 0 & 0 \\
 -\frac{2}{9} & -\frac{2}{9} & \frac{4}{9} & \frac{4}{9} & -\frac{2}{9} & -\frac{2}{9} & 0 & 0 & 0 & 0 & 0 & 0 & 0 & 0 & 0 & 0 & 0 & 0 & 0 & 0 & 0 \\
 -\frac{2}{9} & -\frac{2}{9} & \frac{4}{9} & \frac{4}{9} & -\frac{2}{9} & -\frac{2}{9} & 0 & 0 & 0 & 0 & 0 & 0 & 0 & 0 & 0 & 0 & 0 & 0 & 0 & 0 & 0 \\
 -\frac{2}{9} & \frac{4}{9} & -\frac{2}{9} & -\frac{2}{9} & \frac{4}{9} & -\frac{2}{9} & 0 & 0 & 0 & 0 & 0 & 0 & 0 & 0 & 0 & 0 & 0 & 0 & 0 & 0 & 0 \\
 \frac{4}{9} & -\frac{2}{9} & -\frac{2}{9} & -\frac{2}{9} & -\frac{2}{9} & \frac{4}{9} & 0 & 0 & 0 & 0 & 0 & 0 & 0 & 0 & 0 & 0 & 0 & 0 & 0 & 0 & 0 \\
 0 & 0 & 0 & 0 & 0 & 0 & \frac{19}{28} & \frac{5}{56} & \frac{5}{56} & \frac{5}{56} & 0 & -\frac{5}{56} & \frac{5}{56} & 0 & -\frac{5}{56} & -\frac{5}{56} & -\frac{5}{56} &
   -\frac{19}{28} & 0 & 0 & 0 \\
 0 & 0 & 0 & 0 & 0 & 0 & \frac{5}{56} & \frac{19}{28} & \frac{5}{56} & 0 & \frac{5}{56} & -\frac{5}{56} & -\frac{5}{56} & -\frac{5}{56} & -\frac{19}{28} & \frac{5}{56} & 0 &
   -\frac{5}{56} & 0 & 0 & 0 \\
 0 & 0 & 0 & 0 & 0 & 0 & \frac{5}{56} & \frac{5}{56} & \frac{19}{28} & -\frac{5}{56} & -\frac{5}{56} & -\frac{19}{28} & 0 & \frac{5}{56} & -\frac{5}{56} & 0 & \frac{5}{56} &
   -\frac{5}{56} & 0 & 0 & 0 \\
 0 & 0 & 0 & 0 & 0 & 0 & \frac{5}{56} & 0 & -\frac{5}{56} & \frac{19}{28} & \frac{5}{56} & \frac{5}{56} & \frac{5}{56} & -\frac{5}{56} & 0 & -\frac{5}{56} & -\frac{19}{28} &
   -\frac{5}{56} & 0 & 0 & 0 \\
 0 & 0 & 0 & 0 & 0 & 0 & 0 & \frac{5}{56} & -\frac{5}{56} & \frac{5}{56} & \frac{19}{28} & \frac{5}{56} & -\frac{5}{56} & -\frac{19}{28} & -\frac{5}{56} & \frac{5}{56} &
   -\frac{5}{56} & 0 & 0 & 0 & 0 \\
 0 & 0 & 0 & 0 & 0 & 0 & -\frac{5}{56} & -\frac{5}{56} & -\frac{19}{28} & \frac{5}{56} & \frac{5}{56} & \frac{19}{28} & 0 & -\frac{5}{56} & \frac{5}{56} & 0 & -\frac{5}{56} &
   \frac{5}{56} & 0 & 0 & 0 \\
 0 & 0 & 0 & 0 & 0 & 0 & \frac{5}{56} & -\frac{5}{56} & 0 & \frac{5}{56} & -\frac{5}{56} & 0 & \frac{19}{28} & \frac{5}{56} & \frac{5}{56} & -\frac{19}{28} & -\frac{5}{56} &
   -\frac{5}{56} & 0 & 0 & 0 \\
 0 & 0 & 0 & 0 & 0 & 0 & 0 & -\frac{5}{56} & \frac{5}{56} & -\frac{5}{56} & -\frac{19}{28} & -\frac{5}{56} & \frac{5}{56} & \frac{19}{28} & \frac{5}{56} & -\frac{5}{56} &
   \frac{5}{56} & 0 & 0 & 0 & 0 \\
 0 & 0 & 0 & 0 & 0 & 0 & -\frac{5}{56} & -\frac{19}{28} & -\frac{5}{56} & 0 & -\frac{5}{56} & \frac{5}{56} & \frac{5}{56} & \frac{5}{56} & \frac{19}{28} & -\frac{5}{56} & 0 &
   \frac{5}{56} & 0 & 0 & 0 \\
 0 & 0 & 0 & 0 & 0 & 0 & -\frac{5}{56} & \frac{5}{56} & 0 & -\frac{5}{56} & \frac{5}{56} & 0 & -\frac{19}{28} & -\frac{5}{56} & -\frac{5}{56} & \frac{19}{28} & \frac{5}{56} &
   \frac{5}{56} & 0 & 0 & 0 \\
 0 & 0 & 0 & 0 & 0 & 0 & -\frac{5}{56} & 0 & \frac{5}{56} & -\frac{19}{28} & -\frac{5}{56} & -\frac{5}{56} & -\frac{5}{56} & \frac{5}{56} & 0 & \frac{5}{56} & \frac{19}{28} &
   \frac{5}{56} & 0 & 0 & 0 \\
 0 & 0 & 0 & 0 & 0 & 0 & -\frac{19}{28} & -\frac{5}{56} & -\frac{5}{56} & -\frac{5}{56} & 0 & \frac{5}{56} & -\frac{5}{56} & 0 & \frac{5}{56} & \frac{5}{56} & \frac{5}{56} &
   \frac{19}{28} & 0 & 0 & 0 \\
 0 & 0 & 0 & 0 & 0 & 0 & 0 & 0 & 0 & 0 & 0 & 0 & 0 & 0 & 0 & 0 & 0 & 0 & 0 & 0 & 0 \\
 0 & 0 & 0 & 0 & 0 & 0 & 0 & 0 & 0 & 0 & 0 & 0 & 0 & 0 & 0 & 0 & 0 & 0 & 0 & 0 & 0 \\
 0 & 0 & 0 & 0 & 0 & 0 & 0 & 0 & 0 & 0 & 0 & 0 & 0 & 0 & 0 & 0 & 0 & 0 & 0 & 0 & 0 \\
\end{array}
\right)}\, .
\ea
 Note that ${\cal M}$ does include from the left and right projectors onto the space spanned by the $\chi^+$-vectors, which are orthogonal to the length metric vectors.

The explicit expression for the lowest order of the correction $K^{\chi^+}=-{\cal R}^\dagger \cdot {\cal M}\cdot {\cal R}$ can be easily computed \cite{LinkNB}, but is not particularly insightful.  The structure of ${\cal R}$ and ${\cal M}$ does provide us however with some essential information about $K^{\chi^+}$. Firstly (for general momenta), ${\cal R}$ has exactly four (right) null vectors, which describe linearized diffeomorphisms, and thus coincide with the null modes of the linearized Einstein-Hilbert Lagrangian.  Indeed, the Area Regge action linearized on a flat  background satisfying the area-length constraints, is invariant under linearized diffeomorphisms \cite{ADH1}.  ${\cal R}$ is thus of rank 6 and hence has also 12 left null vectors. 

These diffeomorphism null vectors for ${\cal R}$ are inherited as null vectors for the correction $K^{\chi^+}$. But  $K^{\chi^+}$ has an additional null vector, given by $n_I= \delta_{I_1 I_2}$.  This null vector arises because of the projectors $\chi^+_+$ included in ${\cal M}$. That is $\chi^+_+ \cdot {\cal R}$ has five null vectors, given by the linearized diffeomorphisms and $n_I$.  The latter null vector  imposes  trace freeness, whereas the linearized diffeomorphisms span the longitudinal modes.  Thus the null modes do agree with the null vectors of the spin-2 projector, which projects onto transversal and trace-free modes.

The  correction can thus also be written in the form $K^{\chi^+}={}^2\!\Pi_{\rm L} \cdot  {\cal K}  \cdot  {}^2\!\Pi_{\rm L}$ where ${\cal K}$ is a $10\times 10$ matrix of order $\Lambda^4$. One choice for ${\cal K}$ is ${\cal K}=-{\cal R}^\dagger\cdot {\cal M} \cdot {\cal R}$, but there might be choices, which lead to less involved matrix elements.

\subsection{Geometric interpretation of the correction term in terms of Weyl curvature}\label{Sec:Weyl}

The Area Regge Hessian is given by
\ba
H_{tt'} =
 \frac{1}{2A_t}  \frac{\partial \epsilon^A_t}{\partial A^2_{t'}} 
 \ea
  that is the rows define (rescaled) linearized deficit angles in the area square variables.  In Length Regge calculus the deficit angles measure curvature and satisfy Bianchi identities, see e.g. \cite{BarrettWilliams}.  These identities are violated in Area Regge calculus \cite{ARE1}, the deficit angles in Area Regge calculus therefore represent rather some conglomerate of curvature and shape mismatching or, as suggested in \cite{DittrichRyan,BFCG2,ARE1}, torsion. 

Contracting the Area Regge Hessian with the area -- length Jacobian $\partial A_t^2/\partial L^2_e$ we obtain the (rescaled) linearized deficit angles as function of the length square fluctuations
\be
\tilde \varepsilon_t = \frac{1}{2A_t}  \frac{\partial \epsilon^L_t}{\partial L^2_e}  \q .
\ee
Note that by acting with the Jacobian $\partial A_t^2/\partial L^2_e$, we also project onto the subspace of area fluctations induced by length fluctuations. This leads to linear dependencies between the linearized deficit angles, which amount to the Bianchi identities \cite{BarrettWilliams}. These Bianchi identities can be used to characterize the area -- length constraints \cite{ARE1}.

The matrix ${\cal R}$  arises basically from the contraction of the lattice Hessian with the  area -- length Jacobian. (Additionally we transform the length fluctuations to length metric fluctuations on the right side, and furthermore consider only the Plus sector of the area fluctuations, which are also transformed to area metric variables.)  One could thus conjecture that ${\cal R}$ does describe some linearized curvature tensor. There is a potential caveat in this interpretation: our lattice Hessian is built from the effective Hessians $H^{\bf C'}$ for the hypercubes, where we integrated out the area fluctuations associated to the bulk triangles. This effective Hessian does therefore already include terms resulting from the coupling of length degrees of freedom and the area-length constraints. We have seen that these terms disappear if we contract the effective Hessian from both sides with the length metric vectors, as this re-produces the Hessian describing the linearized Einstein-Hilbert action. Such terms might still be present, if we contract the effective Hessian only from one side with the length metric vectors, as we do for the computation of ${\cal R}$.

This worry is not confirmed: it turns out that the rows of the matrix ${\cal R}$ (at leading order in $\Lambda$) do describe the components of the Riemann tensor, linearized on a flat background. (As we integrated out areas associated to the bulk of the hypercubes we have only 18 rows, and thus find only 18 of the 20 independent  components of the Riemann tensor: the components $R_{0123},R_{0231},R_{0312}$, which satisfy the cyclicity condition $R_{0123}+R_{0231}+R_{0312}$, are missing.) However the label $B=(ij)(kl)$ does not give the component $R_{ijkl}$ directly, we rather have to dualize\footnote{This need for dualization can be understood from the fact that the deficit angles associated to a given triangle describe the rotation angle resulting from a  parallel transport of a vector around this triangle and in a plane {\it orthogonal} to this triangle.} the pairs $(ij)$ and $(kl)$. Explicitly, the Riemann tensor linearized on a flat background, and in Cartesian coordinates, is given as 
\ba\label{LRiem}
\delta R^{opqr} &=&\tfrac{1}{2}
(\delta^{pm} \delta^{rn} \delta^{os} \delta^{qt} + \delta^{om}\delta^{qn} \delta^{ps} \delta^{rt} - \delta^{pm} \delta^{qn} \delta^{os} \delta^{rt}- \delta^{om} \delta^{rn} \delta^{ps} \delta^{qt}) \,k_mk_n \,\, \delta g_{st}
 \nn\\
&=&\tfrac{1}{4}  \,\{\tfrac{1}{4}{\epsilon^{op}}_ {ij} {\epsilon^{qr}}_{ kl}\} \,\, \left[ ( \epsilon^{ij sm} \epsilon^{kl tn} +\epsilon^{ij tm} \epsilon^{kl sn}) \,k_mk_n \right]\, \,\delta g_{st} \q ,
\ea
where in (\ref{LRiem}), and for the  remaining equations in this section we apply Einstein's summation convention  and $\delta g_{st}$ denotes the length metric fluctuation. (We omitted factors of the determinant of the background metric, as they are equal to unity.)
The term in curly brackets is the dualization map and the term in square brackets in the second line agrees (modulo global factors) with the expression (\ref{3.14})  for the components of ${\cal R}$. (The factors ${\cal S}(B)$ and ${\cal S}(I)$ do not appear in (\ref{LRiem}) as  the summation convention in our matrix notation is different from the one used in (\ref{3.14}).)

The correction $K^{\chi^+}$ is  build from a contraction of ${\cal R}$ with itself, using  a `super metric' ${\cal M}$. But ${\cal M}$ includes projectors $\chi^+_+$ on both sides, thus $K^{\chi^+}$ is actually a contraction of $\chi^+_+\cdot {\cal R}$, again with a `super metric' ${\cal M}$.  The rows of $\chi^+_+\cdot {\cal R}$ do give us (a multiple of) components of the linearized Weyl tensor. Again we have to dualize the index pairs in $B$, to get the corresponding index combination for the Weyl tensor. 

To see this, we note first that this dualization commutes with $\chi^+_+$. More concretely we have
\ba\label{Dualchi}
\{\tfrac{1}{4}{\epsilon^{op}}_ {ij} {\epsilon^{qr}}_{ kl}\} ({}^t \chi^+_+)_{i'j'k'l'}^{ijkl} \,=\, ({}^t \chi^+_+)^{opqr}_{ijkl}  \{\tfrac{1}{4}{\epsilon^{ij}}_ {i'j'} {\epsilon^{kl}}_{ k'l'}\}  \,=\,  ({}^t \chi^+_+)_{i'j'k'l'}^{opqr}
\ea
where ${}^t \chi^+_+$ denotes the projector $\chi^+_+$ in tensor notation, see Appendix \ref{AppD}.  That is $\chi^+_+$ is invariant under the dualization map.  Applying $\chi^+_+$ to the linearized $\delta R^{opqr}$ projects out all possible traces and leads to the linearized Weyl curvature
\ba
\delta C^{opqr} &=&  ({}^t \chi^+_+)^{opqr}_{o'p'q'r'} \, \, \delta R^{ o'p'q'r'}        \q .
\ea
Note, that although we are missing the components $B=(01)(23),(02)(31),(03)(12)$ from ${\cal R}$, this does not affect our conclusion, as the dualization map only permutes these components among themselves and the operation of projecting out all traces does not affect these  components.

In summary, the $\chi^+$ induced correction to the linearized Einstein Hilbert action is a contraction of the linearized Weyl curvature with itself. But this contraction employs a (degenerate) `super metric' ${\cal M}$, that is\footnote{Note that ${\cal M}$ is invariant under dualization of its left or right indices. The reason is, that it does contain from both sides the projector $\chi^+_+$, which satisfies (\ref{Dualchi}). }
\ba
 \frac{1}{2} (\delta g)^\dagger \cdot (K^{\chi^+})\cdot \delta g   \, \,=\, \, \,\, \frac{\Lambda^4}{\lambda^2} \,\,  \delta C^{opqr} (-k) \,\, (\tfrac{1}{\lambda^6}  \,{}^t\!{\cal M}  )_{opqro'p'q'r'} \,\, \delta C^{o'p'q'r'} (k) \, + {\cal O}(\Lambda^5)
\ea
where $(\lambda^{-6} )\, \,{}^t\!{\cal M}$ is ${\cal M}$ in tensor notation ( see Appendix \ref{AppD}), and made dimensionless.

%We could add to ${}^t\!{\cal M}$ terms that will be projected out by $\chi^+_+$  ---  this would not change the result if we use the linearized Weyl curvature  for the contraction. This freedom is however not sufficient to turn ${}^t{\cal M}$ into a diagonal form. The most we can achieve, is to bring the block for the diagonal area metric components (i.e. for the indices $B=(01)(01),\ldots$) to diagonal form, and to e.g. eliminate all negative entries in the block for the non-diagonal area metric components of the form $B=(01)(0,2), \ldots, (31)(32)$.

The tensor  ${}^t\!{\cal M}$ is, different from the tensors we considered so far in this section,  not covariant. That is, it cannot be built from contractions of Kronecker Deltas and Levi-Civita densities. Indeed one can show, that ${}^t\!{\cal M}$ is {\it not} invariant under coordinate transformations that amount to general rotations in the flat background geometry. But a tensor constructed from contraction of Deltas and Levi-Civita symbols is invariant under such rotations.  ${}^t\!{\cal M}$ is however invariant under rotations that leave the hypercubical lattice invariant (but might change the directions of the diagonal edges in the triangulations). That is, simple rotations in planes spanned by two lattice vectors with rotation angle $\alpha= \pi/2$ and their combinations.

The invariance of the action under such hypercubical symmetries is likely to play an important role  in the restoration of full rotation invariance under coarse graining \cite{Improved,BahrDittrichHe}.

Furthermore, we note that the numerical values of the components of ${\cal M}$ are approximating the numerical values of a multiple of the projector ${\chi}^+_+$, if we exclude in our considerations the components $B=(01)(23),(02)(31),(03)(12)$. Rescaling ${\cal M}$, so that we match the $B=(01)(01), \ldots$ components with those in ${\chi}^+_+$, we can compare the remaining entries: here we find $0.509$ in ${\cal M}$ instead of $0.5$ in ${\chi}^+_+$ and $0.067$ in ${\cal M}$ instead of $0$ in ${\chi}^+_+$.

One source for the non-covariance of ${}^t\!{\cal M}$ might be our procedure, where we integrated out the bulk triangles from the hypercubes. We thus are missing the components $B=(01)(23),(02)(31),(03)(12)$ of the area metric fluctuations. We leave to future work the exploration of the possibility to extract from the bulk triangles such components, and to integrate out only the remaining bulk degrees of freedom.

~\\
To summarize, the effective action for the length fluctuations to the two lowest orders in the lattice constant, encoded in Area Regge calculus, can be understood to arise from an effective action for area metric fluctuations. The area metric fluctuations can be split into a trace part, which can be re-organized into length metric fluctuations, and a trace-less part. This is analogous to the splitting of the Riemann tensor into  Ricci tensor on the one hand, and Weyl tensor on the other. 

The length metric fluctuations are mass-less, whereas the trace-less parts of the area metric are massive.  The part of the action quadratic in the length metric fluctuations is given by a lattice version of the linearized Einstein-Hilbert action. The coupling between the trace parts and the trace-less parts is described by the linearized Weyl tensor. To describe the two lowest orders in the effective dynamics of the length metric fluctuations, we then just need the mass term for the trace-less parts of the area metric. This mass term is described by a non-covariant tensor, which is {\it not} invariant under general rotations, but is invariant under lattice preserving rotations.

\section{Summary and discussion}\label{Disc}

The Area Regge action arises in the semi-classical limit of spin foam models \cite{SFLimit} and hence is of fundamental importance for a better understanding of the dynamics encoded in loop quantum gravity and spin foams.  But, until recently, there have been only few works investigating the dynamics defined by this action \cite{Wainwright,ADH1}. 

Moreover, due to a misleading identification of the deficit angle appearing in Length Regge calculus with the  deficit angle in Area Regge Calculus, it has been widely thought that the equations of motions for Area Regge calculus demand flatness, and can therefore not lead to general relativity.  But the deficit angle in Area Regge calculus is a much more involved object than in Length Regge Calculus: whereas the deficit angle in Length Regge Calculus measures curvature in a torsion free space-time, the deficit angle in Area Regge Calculus rather conglomerates curvature and torsion. The latter appears as a shape mismatching of tetrahedra and triangles \cite{AreaAngle,DittrichRyan,BFCG2}.

The recent work \cite{ARE1} considered linearized Area Regge calculus on a tilted version of the standard triangulation of the hypercubic lattice, and showed that, contrary to expectations, the linearized Einstein-Hilbert action emerges in the continuum limit. That is, it constitutes the leading contribution, namely at order ${\cal O}(\lambda^0)$  in the lattice constant $\lambda$.  Area Regge calculus on the standard triangulation of the hypercubic lattice is however singular -- this is why the tilting has been introduced in \cite{ARE1}. In the limit where the tilting parameter goes to zero, one obtains so-called hypercubical constraints, that suppress a considerably number of the degrees of freedom. This includes degrees of freedom, that one could interpret as trace free parts of an area metric.  The trace parts of the area metric can be organized into a length metric, whose dynamics is described by the linearized Einstein-Hilbert action. There are yet additional degrees of freedom (corresponding to the Minus sector defined in Section \ref{Sec:separation}), which come however with a mass proportional to the inverse of the lattice constant. Integrating out these degrees of freedom leads to a correction terms of order ${\cal O}(\lambda^4)$ in the lattice constant.

One could however wonder whether it is only due to the hypercubical constraints, that one obtains in leading order the Einstein-Hilbert action, and thus whether the results of \cite{ARE1} are restricted to the standard triangulation of the hypercubic lattice. 

We therefore considered in this work a triangulation on which the Area Regge action is not singular, and which does thus not feature hypercubical constraints. This new triangulation is a simple variant of the standard triangulation, obtained by inserting into each hypercube a central vertex. Although this new triangulation has double as many area variables per vertex than the standard triangulation (100 versus 50), it leads --- after integrating out the 64 bulk triangles in each hypercube -- to a number of simplifications as compared to \cite{ARE1}. 

This new triangulation does also lead, at the lowest order ${\cal O}(\lambda^0)$ in the lattice constant, to the linearized Einstein-Hilbert action. This is due to the fact that almost all  degrees of freedom come with a mass term, which is inversely proportional to the lattice constant. The exceptions are 10 degrees of freedom per vertex, which can be used to define a length metric.  We can thus assume that this result is robust, and does, in particular not depend on choosing a triangulation for which the Area Regge action is singular. 

The most consequential change, when going from the standard triangulation of the hypercubic lattice to its centrally subdivided triangulation, is that there are no hypercubical constraints, suppressing the trace free parts of the area metric.  These degrees of freedom come from the Plus sector (defined in Section \ref{Sec:separation}) and lead, if integrated out, to the lowest order correction on top of the linearized Einstein-Hilbert action.\footnote{Note that the lattice version (\ref{LEHL}) of the linearized Einstein-Hilbert action does also include ${\cal O}(\lambda^1)$ and ${\cal O}(\lambda^2)$ terms, which arise, if one expands the $\omega$-factors into a power series in the momenta.} This correction is now of order ${\cal O}(\lambda^2)$, whereas the Minus sector leads to ${\cal O}(\lambda^4)$ corrections --- as was the case for the standard triangulation. 

The two lowest orders in the effective graviton dynamics of Area Regge Calculus can thus be understood to arise from an effective action for the area metric. We note that as a side result of this work, we show how Area Regge calculus can be used to define an action for area metrics, which re-produces gravitons as the only propagating degrees of freedom.

This  action gives mass to the trace parts of the area metric, but keeps the trace degrees of freedom, which can be re-organized into a length metric, as mass-less. The action contribution from the latter degrees of freedom is given by the linearized Einstein-Hilbert action.  The coupling between the length metric degrees of freedom and the trace-less parts of the area metric is described by the linearized Weyl curvature tensor.  The mass for the trace-less parts of the area metric is described by a tensor, which, unfortunately, is not covariant, but can be approximated by the projector to the trace-less modes of the area metric.

 The dynamics for the length metric fluctuations is thus described by the linearized Einstein-Hilbert action at lowest order $\lambda^0$, and by a quadratic form in the linearized Weyl curvature tensor, at order $\lambda^2$. 

Let us comment on two points here: Firstly, we saw that the (Area) Regge action leads naturally to a reconstruction of  both the linearized Riemann tensor and the linearized Weyl tensor, and not only the Einstein tensor.   This can facilitate the construction of higher order curvature invariants on the lattice, see also \cite{HamberWilliamsHO}, or of curvature observables for Regge calculus and spin foams.
 Secondly, although the tensor which describes the mass terms for the trace-less parts of the area metric fluctuations is not covariant, it is invariant under rotations that are symmetries of the lattice. This indicates that a process, where we refine the lattice and integrate out the additional degrees of freedom \cite{Improved,BahrDittrichHe,PerfectPI}, might lead to an action, which is invariant under the full rotation group, and therefore to a covariant tensor for the mass terms.

This brings us to a number of possible follow-up directions for research: One could consider even more refined lattices, e.g. where, in addition to the hypercubes, one also subdivides the three-dimensional cubes with vertices, or even the cubes and the squares. There will be a further increase in the number of area variables per vertex one has to deal with. But  a range of technical improvements introduced in the course of this work and accessible via the accompanying notebook \cite{LinkNB}, might make this task possible.  The question would be whether these lattices would lead to a fully covariant correction term at second order in the lattice constant. More generally one could consider a systematic coarse graining flow \cite{Improved,BahrDittrichHe}, that would connect triangulations of the same type, but with different lattice constants. 

To understand better the meaning and dynamics of the degrees of freedom of Area Regge caculus and eventually spin foams, we could also perform a canonical analysis of the various lattice actions \cite{DittrichHoehn1}, which would entail an analysis of the constraints encoded in the action. The Area Regge action is also available in a first order version \cite{ADH1}, where one uses areas and (four-dimensional) dihedral angles as variables. Can these be used to reconstruct a lattice connection from which one can derive lattice expressions for torsion and curvature?

An alternative approach is to consider continuum actions, that can describe the  Area Regge dynamics in the continuum limit. Possible starting points are $BF$-like actions with potential terms for the $B$-fields \cite{Krasnov} or geometrical natural actions for the area metric \cite{Schuller1,Schuller2}.

Let us return to the relevance of our findings for the continuum limit of spin foams. This work, along with \cite{ARE1}, gives a new and rather unexpected\footnote{A first hint appeared in the review \cite{ReggeWilliams}, but sadly, it  seemed not to have followed up at that time.  The authors discuss in a short section, that the linearized Area Regge action on a tilted version of the standard triangulation, does lead to only 10 massless degrees of freedom per vertex, despite having 50 degrees of freedom per vertex. They comment that the analytical evaluation of this action was not possible (at that time) however. We assume that this means that one could not keep the $\omega_i$ factors general, but had to insert explicit numerical choices for these factors. To determine whether degrees of freedom come with a mass term it is sufficient to set all the $\omega$ factors equal to $1$.} mechanism, that resolves the flatness problem of spin foams in the continuum limit. This mechanism is different from arguments in \cite{MuxinSmallBI, EffSF1,EffSF2,ComplexSP}, and the mechanism underlying the explicit numerical evaluations \cite{EffSF1,EffSF2,EffSF3} for triangulations with few simplices.  In contrast to these arguments the mechanism here  does  not impose a bound on the Barbero-Immirzi parameter, which controls how strongly the  constraints are imposed, that restrict the area variables to arise from a consistent length assignment \cite{EffSF1}. In fact, it also works if we do not impose at all such constraints, and might  therefore ensure that the Barrett-Crane model \cite{BC} features gravitons in its continuum limit. But for this mechanism to come into play, a large lattice with many building blocks is key, as the relevant scaling only emerges for sufficiently large lattices. 

We furthermore showed explicitly, that the (linearized) Einstein-Hilbert action does emerge to leading order in the lattice constant, and we thus have propagating gravitons. We also find an explicit expressions for the next order correction for the graviton dynamics, the corresponding action term is quadratic in the Weyl curvature tensor. It will be interesting to understand the phenomenological implications of this correction term.

Both the Barrett-Crane spin foam model \cite{BC}, as well as the EPRL-FK models \cite{EPRL-FK}  showcase the Area Regge action in their semi-classical limit \cite{SFLimit}. The effective spin foam models \cite{EffSF1,EffSF3} rely even more directly on the  Area Regge action. The results here therefore indicate that these models can lead to propagating gravitons in their continuum limit, and might be universal for spin foam models leading to the area action.  This is despite the arguments of \cite{Alesci} for the Barrett-Crane model, which involved the analysis of the amplitude for a single four-simplex only. 

The EPRL-FK models  and effective spin foam models do impose the area-length constraints weakly, that is include  an explicit mechanism for the suppression of 'pure' area degrees of freedom, that do not arise from length degrees of freedom. This is most transparent for the effective spin foam models \cite{EffSF1,EffSF2,EffSF3}, where this suppression is implemented via an imaginary mass terms for these degrees of freedom. Includding these additional terms does however not change substantially the results found for the Area Regge action in the case of the standard triangulation of the hypercubic lattice \cite{ARE1}:  even the form of the leading order correction remains the same. We expect that this will be also the case for the triangulation considered in this work, but will leave this investigation for future work. 

  The results presented here give a much greater confidence, that general relativity will emerge in the continuum limit of spin foams. We did however rely here on a  expansion around a background field, assuming that the non-perturbative continuum limit of spin foams does allow for such an expansion.\footnote{The construction of representations for loop quantum gravity, which are based on vacua peaked on flat connections \cite{DG}, instead of vacua peaked on geometries with vanishing volumes, make this scenario more likely.} To confirm this, we believe that it will be essential to control two kinds of divergencies \cite{SFDiv} in spin foams:  The first term arises from a possible restoration of diffeomorphism symmetry in the continuum limit \cite{DiffDiv}, and could thus be controlled by appropriately dividing out gauge volumes. Here it will be also essential to construct diffeomorphism invariant measures \cite{ PImeasure1,BahrStein}.  The second kind arises because the integration (or rather summation) range for spin foams are typically infinite, and can include so-called spike configurations \cite{Ambjorn:1997ub}, which are reflections of the conformal factor problem \cite{Gibbons:1978ac}. This has led to often insurmountable issues for the continuum limit in Euclidean approaches to quantum gravity \cite{LollLR}. But spin foams rely on a proper quantum mechanical (non-Wick rotated) path integral. Finding new methods to evaluate such quantum gravitational path integrals with infinite integration or summation ranges might avoid these issues \cite{LorCal}.

\appendix

\section{Construction of projectors with an over-complete basis}\label{AppA}

In this section we will discuss how to construct a projector, which projects onto the space spanned by a possibly over-complete set ${\cal V}=\{v_A\}_{A=1}^n$ of vectors. One possibility is to orthonormalize this set of vectors  and build the projectors from a sum of one-dimensional ones. Such an orthonormalization procedure can however be very cumbersome. 

An alternative method proceeds by defining the Gram matrix 
\ba
G_{AA'} &=& v_A^\dagger \cdot v_{A'}     \q .
\ea
If the set ${\cal V}$ is over-complete, the Gram matrix will have a set of null vectors $\{n_b\}_b$, which satisfy $\sum_{A'}G_{AA'}(n_b)_{A'}=0$.  In this case we form the matrix
\ba
\tilde G_{AA'} &=&  G_{AA'} + \sum_b \beta_b  \, (n_b)_A (n_b^\dagger)_{A'} \q, 
\ea
which, for generic parameters $\beta_b$, can be inverted.  We can then define the projector 
\ba
\Pi&=&       \sum_{A,A'}      v_A  \,  ( \tilde G^{-1})_{AA'}  \, v_{A'}^\dagger       \q ,
\ea
which is independent of the choice for the $\beta_b$ and satisfies
\ba
\Pi \cdot v_B &=&=   \sum_{A,A'}      v_A  \,  ( \tilde G^{-1})_{AA'}  \, G_{A'B} \,\,=\,\, v_B
\ea
for all $A=1,\ldots, n$, by construction. We have also $\Pi \cdot w=0$ for any vector $w$, which is orthogonal to the set ${\cal V}$, that is, which satisfies $v_A^\dagger \cdot w=0$.

\section{Transformation from area squares to area metric and from length square to length metric variables}\label{AppB}

In Section \ref{Sec:separation} we discussed the following transformation between the (Plus sector) area square fluctuations and the area metric fluctuations:
\ba\label{AMAtrafo3App}
\alpha^+_{(2^i,2^j)} &=& {}^A\!\mu^+_{(ij)(ij)}          \q \text{with}\,\,\, i<j \nn\\
\alpha^+_{(2^i,2^j+2^k)}&=& {}^A\!\mu^+_{O(ij) O(ij)}+ {}^A\!\mu^+_{O(ik) O(ik)} +  2{}^A\!\mu^+_{(ij) (ik)}      \q \text{with}\,\,i\neq j,k \, \text{and}\,  j<k \q . 
\ea
We can summarize this transformation into a matrix ${}^A\! M$, so that $\alpha^+= {}^A\! M \cdot {}^A\!\mu^+$. This matrix is given by
\ba
{}^A\! M=
{\tiny\left(
\begin{array}{cccccccccccccccccc}
 1 & 0 & 0 & 0 & 0 & 0 & 0 & 0 & 0 & 0 & 0 & 0 & 0 & 0 & 0 & 0 & 0 & 0 \\
 0 & 1 & 0 & 0 & 0 & 0 & 0 & 0 & 0 & 0 & 0 & 0 & 0 & 0 & 0 & 0 & 0 & 0 \\
 0 & 0 & 1 & 0 & 0 & 0 & 0 & 0 & 0 & 0 & 0 & 0 & 0 & 0 & 0 & 0 & 0 & 0 \\
 0 & 0 & 0 & 1 & 0 & 0 & 0 & 0 & 0 & 0 & 0 & 0 & 0 & 0 & 0 & 0 & 0 & 0 \\
 0 & 0 & 0 & 0 & 1 & 0 & 0 & 0 & 0 & 0 & 0 & 0 & 0 & 0 & 0 & 0 & 0 & 0 \\
 0 & 0 & 0 & 0 & 0 & 1 & 0 & 0 & 0 & 0 & 0 & 0 & 0 & 0 & 0 & 0 & 0 & 0 \\
 1 & 1 & 0 & 0 & 0 & 0 & 2 & 0 & 0 & 0 & 0 & 0 & 0 & 0 & 0 & 0 & 0 & 0 \\
 1 & 0 & 1 & 0 & 0 & 0 & 0 & 2 & 0 & 0 & 0 & 0 & 0 & 0 & 0 & 0 & 0 & 0 \\
 0 & 1 & 1 & 0 & 0 & 0 & 0 & 0 & 2 & 0 & 0 & 0 & 0 & 0 & 0 & 0 & 0 & 0 \\
 1 & 0 & 0 & 1 & 0 & 0 & 0 & 0 & 0 & 2 & 0 & 0 & 0 & 0 & 0 & 0 & 0 & 0 \\
 1 & 0 & 0 & 0 & 1 & 0 & 0 & 0 & 0 & 0 & 2 & 0 & 0 & 0 & 0 & 0 & 0 & 0 \\
 0 & 0 & 0 & 1 & 1 & 0 & 0 & 0 & 0 & 0 & 0 & 2 & 0 & 0 & 0 & 0 & 0 & 0 \\
 0 & 1 & 0 & 1 & 0 & 0 & 0 & 0 & 0 & 0 & 0 & 0 & 2 & 0 & 0 & 0 & 0 & 0 \\
 0 & 1 & 0 & 0 & 0 & 1 & 0 & 0 & 0 & 0 & 0 & 0 & 0 & 2 & 0 & 0 & 0 & 0 \\
 0 & 0 & 0 & 1 & 0 & 1 & 0 & 0 & 0 & 0 & 0 & 0 & 0 & 0 & 2 & 0 & 0 & 0 \\
 0 & 0 & 1 & 0 & 1 & 0 & 0 & 0 & 0 & 0 & 0 & 0 & 0 & 0 & 0 & 2 & 0 & 0 \\
 0 & 0 & 1 & 0 & 0 & 1 & 0 & 0 & 0 & 0 & 0 & 0 & 0 & 0 & 0 & 0 & 2 & 0 \\
 0 & 0 & 0 & 0 & 1 & 1 & 0 & 0 & 0 & 0 & 0 & 0 & 0 & 0 & 0 & 0 & 0 & 2 \\
\end{array}
\right)
}
\ea
where the rows are labelled by triangle pairs, which we denoted by (see explanation below Equation (\ref{trafo2}))
\ba
&&(1,2),(1,4),(1,8),(2,4)(2,8),(4,8), \nn\\
&&(1,6),(1,10), (1,12),(2,5), (2,9),(2,10), (4,3),(4,9), (4,10),(8,3),(8,5),(8,6)\, ;
\ea
and the columns label independent area metric components:
\ba\label{AppBIndex}
B=&&(01)(01),(02)(02),(03)(03),(12)(12),(13)(13),(23)(23), \nn\\
&&(01)(02), (02)(03), (01)(03), 
(10)(12), (10)(13),(12)(13), \nn\\
&&(20)(21), (20)(23), (21)(23), 
(30)(31), (30)(32), (31)(32) \, .
\ea

Similarly, we have a transformation matrix  ${}^L\! M$ between the length square fluctuations and the length metric fluctuations
\ba
{}^L\! M=
{\tiny
\left(
\begin{array}{cccccccccc}
 1 & 0 & 0 & 0 & 0 & 0 & 0 & 0 & 0 & 0 \\
 0 & 1 & 0 & 0 & 0 & 0 & 0 & 0 & 0 & 0 \\
 0 & 0 & 1 & 0 & 0 & 0 & 0 & 0 & 0 & 0 \\
 0 & 0 & 0 & 1 & 0 & 0 & 0 & 0 & 0 & 0 \\
 1 & 1 & 0 & 0 & 2 & 0 & 0 & 0 & 0 & 0 \\
 1 & 0 & 1 & 0 & 0 & 2 & 0 & 0 & 0 & 0 \\
 0 & 1 & 1 & 0 & 0 & 0 & 0 & 2 & 0 & 0 \\
 1 & 0 & 0 & 1 & 0 & 0 & 2 & 0 & 0 & 0 \\
 0 & 1 & 0 & 1 & 0 & 0 & 0 & 0 & 2 & 0 \\
 0 & 0 & 1 & 1 & 0 & 0 & 0 & 0 & 0 & 2 \\
\end{array}
\right)
}\q .
\ea
The rows are labelled by the edges 
\ba
\{0,X\}&=& \{0,1\},\{0,2\},\{0,4\},\{0,8\},\nn\\
&&\{0,3\},\{0,5\},\{0,9\},\{0,6\},\{0,10\},\{0,12\}
\ea
and the columns by the independent components of the length metric
\ba
I=&&00,11,22,33,01,02,03,12,13,23\, .
\ea

\section{Projectors $\chi^+_+$ and $\chi^-_-$}\label{AppC}

We defined projectors $\chi^+_+$ and $\chi^-_-$ in Section \ref{Sec:Length}, which project out area metric fluctuations that arise from length square fluctuations, projected to the Plus and Minus sector, respectively.  Rows and columns of these projectors are labelled by the index $B$, whose values we detailed in (\ref{AppBIndex}).

The projector $\chi^+_+$ has matrix rank 8, and is given by 
\ba
\chi^+_+=\frac{1}{6}
{\tiny
\left(
\begin{array}{cccccccccccccccccc}
 2 & -1 & -1 & -1 & -1 & 2 & 0 & 0 & 0 & 0 & 0 & 0 & 0 & 0 & 0 & 0 & 0 & 0 \\
 -1 & 2 & -1 & -1 & 2 & -1 & 0 & 0 & 0 & 0 & 0 & 0 & 0 & 0 & 0 & 0 & 0 & 0 \\
 -1 & -1 & 2 & 2 & -1 & -1 & 0 & 0 & 0 & 0 & 0 & 0 & 0 & 0 & 0 & 0 & 0 & 0 \\
 -1 & -1 & 2 & 2 & -1 & -1 & 0 & 0 & 0 & 0 & 0 & 0 & 0 & 0 & 0 & 0 & 0 & 0 \\
 -1 & 2 & -1 & -1 & 2 & -1 & 0 & 0 & 0 & 0 & 0 & 0 & 0 & 0 & 0 & 0 & 0 & 0 \\
 2 & -1 & -1 & -1 & -1 & 2 & 0 & 0 & 0 & 0 & 0 & 0 & 0 & 0 & 0 & 0 & 0 & 0 \\
 0 & 0 & 0 & 0 & 0 & 0 & 3 & 0 & 0 & 0 & 0 & 0 & 0 & 0 & 0 & 0 & 0 & -3 \\
 0 & 0 & 0 & 0 & 0 & 0 & 0 & 3 & 0 & 0 & 0 & 0 & 0 & 0 & -3 & 0 & 0 & 0 \\
 0 & 0 & 0 & 0 & 0 & 0 & 0 & 0 & 3 & 0 & 0 & -3 & 0 & 0 & 0 & 0 & 0 & 0 \\
 0 & 0 & 0 & 0 & 0 & 0 & 0 & 0 & 0 & 3 & 0 & 0 & 0 & 0 & 0 & 0 & -3 & 0 \\
 0 & 0 & 0 & 0 & 0 & 0 & 0 & 0 & 0 & 0 & 3 & 0 & 0 & -3 & 0 & 0 & 0 & 0 \\
 0 & 0 & 0 & 0 & 0 & 0 & 0 & 0 & -3 & 0 & 0 & 3 & 0 & 0 & 0 & 0 & 0 & 0 \\
 0 & 0 & 0 & 0 & 0 & 0 & 0 & 0 & 0 & 0 & 0 & 0 & 3 & 0 & 0 & -3 & 0 & 0 \\
 0 & 0 & 0 & 0 & 0 & 0 & 0 & 0 & 0 & 0 & -3 & 0 & 0 & 3 & 0 & 0 & 0 & 0 \\
 0 & 0 & 0 & 0 & 0 & 0 & 0 & -3 & 0 & 0 & 0 & 0 & 0 & 0 & 3 & 0 & 0 & 0 \\
 0 & 0 & 0 & 0 & 0 & 0 & 0 & 0 & 0 & 0 & 0 & 0 & -3 & 0 & 0 & 3 & 0 & 0 \\
 0 & 0 & 0 & 0 & 0 & 0 & 0 & 0 & 0 & -3 & 0 & 0 & 0 & 0 & 0 & 0 & 3 & 0 \\
 0 & 0 & 0 & 0 & 0 & 0 & -3 & 0 & 0 & 0 & 0 & 0 & 0 & 0 & 0 & 0 & 0 & 3 \\
\end{array}
\right)
} \q .
\ea
With this explicit form at hand, one can check that it projects out all possible traces from the area metric fluctuations. 

The projector $\chi^-_-$ has matrix rank 9 and is given by
\ba
\chi^-_-=\frac{1}{2^4\times 11}
{\tiny
\left(
\begin{array}{cccccccccccccccccc}
 100 & -48 & -52 & 28 & 24 & -4 & 10 & 14 & 0 & 10 & 14 & 0 & -4 & -14 & -14 & 4 & -10 & -10 \\
 -48 & 104 & -56 & -24 & -8 & 16 & 4 & -12 & 0 & 4 & -12 & 0 & 16 & 12 & 12 & -16 & -4 & -4 \\
 -52 & -56 & 108 & -4 & -16 & -12 & -14 & -2 & 0 & -14 & -2 & 0 & -12 & 2 & 2 & 12 & 14 & 14 \\
 28 & -24 & -4 & 124 & -32 & 20 & -6 & -26 & 0 & -6 & -26 & 0 & 20 & 26 & 26 & -20 & 6 & 6 \\
 24 & -8 & -16 & -32 & 136 & -8 & -24 & -16 & 0 & -24 & -16 & 0 & -8 & 16 & 16 & 8 & 24 & 24 \\
 -4 & 16 & -12 & 20 & -8 & 148 & -18 & 10 & 0 & -18 & 10 & 0 & -28 & -10 & -10 & 28 & 18 & 18 \\
 10 & 4 & -14 & -6 & -24 & -18 & 67 & -3 & 0 & -21 & -3 & 0 & -18 & 3 & 3 & 18 & 21 & -67 \\
 14 & -12 & -2 & -26 & -16 & 10 & -3 & 75 & 0 & -3 & -13 & 0 & 10 & 13 & -75 & -10 & 3 & 3 \\
 0 & 0 & 0 & 0 & 0 & 0 & 0 & 0 & 88 & 0 & 0 & -88 & 0 & 0 & 0 & 0 & 0 & 0 \\
 10 & 4 & -14 & -6 & -24 & -18 & -21 & -3 & 0 & 67 & -3 & 0 & -18 & 3 & 3 & 18 & -67 & 21 \\
 14 & -12 & -2 & -26 & -16 & 10 & -3 & -13 & 0 & -3 & 75 & 0 & 10 & -75 & 13 & -10 & 3 & 3 \\
 0 & 0 & 0 & 0 & 0 & 0 & 0 & 0 & -88 & 0 & 0 & 88 & 0 & 0 & 0 & 0 & 0 & 0 \\
 -4 & 16 & -12 & 20 & -8 & -28 & -18 & 10 & 0 & -18 & 10 & 0 & 60 & -10 & -10 & -60 & 18 & 18 \\
 -14 & 12 & 2 & 26 & 16 & -10 & 3 & 13 & 0 & 3 & -75 & 0 & -10 & 75 & -13 & 10 & -3 & -3 \\
 -14 & 12 & 2 & 26 & 16 & -10 & 3 & -75 & 0 & 3 & 13 & 0 & -10 & -13 & 75 & 10 & -3 & -3 \\
 4 & -16 & 12 & -20 & 8 & 28 & 18 & -10 & 0 & 18 & -10 & 0 & -60 & 10 & 10 & 60 & -18 & -18 \\
 -10 & -4 & 14 & 6 & 24 & 18 & 21 & 3 & 0 & -67 & 3 & 0 & 18 & -3 & -3 & -18 & 67 & -21 \\
 -10 & -4 & 14 & 6 & 24 & 18 & -67 & 3 & 0 & 21 & 3 & 0 & 18 & -3 & -3 & -18 & -21 & 67 \\
\end{array}
\right)
}\; .
\ea

\section{ Translating from matrix to tensor notation }\label{AppD}

To describe the dynamics of the area metric fluctuations we have largely used matrices, whose rows and columns are labelled by the independent components of the area metric. An alternative is to use tensor notation, in which the area metric fluctuations appear as $\mu_{ijkl}$, and where one sums over all values for the indices $i,j,k,l=0,1,2,3$.  To translate between these notations, we need to account for the number of times in which an independent area metric component appears in the tensor notation. 

E.g.  the index value $B=(01)(01)$ appears in 4 instances as $(ijkl)=(0101),(1001),(0110),(1010)$ in tensor notation, whereas the index values $B=(01)(02)$ and $B=(01)(23)$ appear in 8 combinations.  Translating  from matrix to tensor notation, we need to divide the components, as given in the matrix notation, by the corresponding factors or by the square root of these factors.  The latter applies if we want to translate the product of two matrices into tensor notation, and want to distribute these factors equally between the two matrices. One might apply the former if one translates the application of a matrix to an area metric fluctuation vector into tensor notation. 

 As an example we detail the projector $\chi^+_+$ in tensor notation. The tensor components $({}^t\! \chi^+_+)_{i'j'k'l'}^{i\,\,j\,\,k\,\,l}$ are given by
\ba
{}^t\! \chi^+_+ = -\tfrac{1}{16} T_1 + \tfrac{1}{12} T_2 +\tfrac{1}{8} T_3
\ea
where 
\ba
(T_1)_{i'j'k'l'}^{i\,\,j\,\,k\,\,l}&=&
\bigg\{ 
\left[
\delta^{ik} \delta_{i'k'} (\delta_{j'}^j \delta_{l'}^l + \delta_{j'}^l \delta_{l'}^j ) \right]  -\left[ i'\leftrightarrow j'\right]-\left[ k'\leftrightarrow l'\right]+\left[ i'\leftrightarrow j, k'\leftrightarrow l'\right]
\bigg\}\nn\\
&&-\{i\leftrightarrow j\} -\{k\leftrightarrow l\}+\{i\leftrightarrow j,k\leftrightarrow l\} \nn \\
(T_2)_{i'j'k'l'}^{i\,\,j\,\,k\,\,l}&=&
(\delta^{ik}\delta^{jl} - \delta^{il} \delta^{jk}) (\delta_{i'k'}\delta^{j'l'} - \delta_{i'l'} \delta_{j'k'})  \nn\\
(T_3)_{i'j'k'l'}^{i\,\,j\,\,k\,\,l}&=&\bigg\{  \left[ \delta_{i'}^i\delta_{j'}^j\delta_{k'}^k\delta_{l'}^l    \right]  -\left[ i'\leftrightarrow j'\right]-\left[ k'\leftrightarrow l'\right]+\left[ i'\leftrightarrow j', k'\leftrightarrow l'\right]\bigg\}\nn\\
&&+\{ i'\leftrightarrow k', j'\leftrightarrow l'\} \q .
\ea
Note that all tensors $T_a,\, a=1,2,3$ satisfy the symmetries of the area metric. Contracting an area metric with $T_1$ amounts to taking the trace on one suitable index pair, e.g. $(ik)$, contracting with $T_2$ amounts to taking the double trace and $T_3$ is proportional to the identity map on the space of area metrics.

\section{ Spin projectors on the lattice}\label{AppE}

Here we will define spin-0, spin-1 and spin-2 projectors on the lattice \cite{BahrDittrichHe}. These projectors are combinations of projectors onto transversal or longitudinal modes, as well as trace modes. They thus involve derivatives.  On the lattice we have the choice between forward and backward derivatives. To allow for both options, we define $\kappa_m=1-\omega_m$ and $\bar{\kappa}_m=1-\bar{\omega}_m$, where $\omega_m:=\exp( \imath\,  \Lambda\, \ k_m),\, m=0,\ldots,3$ are the Fourier transform factors introduced in Section \ref{SecFtrafo}.
The spin projectors are then defined as
\ba\label{disreteprojectors}
({}^0\!\Pi_{\rm L})_{ijkl} &=&\frac{1}{3}
\left(\delta_{ij}+\frac{\bar{\kappa}_i \bar{\kappa}_j}{\Delta'} (1-\delta_{ij} {\kappa}_j) \right)  \,\,
 \left(\delta_{kl}+\frac{\kappa_k \kappa_l}{\Delta'} (1-\delta_{kl}\bar{\kappa}_l) \right)  \,\,
      \nn\\
({}^1\!\Pi_{\rm L})_{ijkl}  &=&   \frac{1}{2}(\delta_{ik}\delta_{jl}+\delta_{il}\delta_{jk})-\nn\\
 &&\frac{1}{2} (1-\delta_{ij}{\kappa}_j)(1-\delta_{kl}\bar \kappa_l)
 \left(
 (\delta_{ik}-\frac{\bar \kappa_i {\kappa}_k}{\Delta'}) \,(\delta_{jl}-\frac{\bar \kappa_j {k}_l}{\Delta'}) +
 (\delta_{il}-\frac{\bar \kappa_i {\kappa}_l}{\Delta'}) \,(\delta_{jk}-\frac{\bar \kappa_j {\kappa}_k}{\Delta'})
 \right)
 \nn\\
({}^2\!\Pi_{\rm L})_{ijkl}&=&
 \frac{1}{2} (1-\delta_{ij}{\kappa}_j)(1-\delta_{kl}\bar \kappa_l)
 \left(
 (\delta_{ik}-\frac{\bar \kappa_i {\kappa}_k}{\Delta'}) \,(\delta_{jl}-\frac{\bar \kappa_j {\kappa}_l}{\Delta'}) +
 (\delta_{il}-\frac{\bar \kappa_i {\kappa}_l}{\Delta'}) \,(\delta_{jk}-\frac{\bar \kappa_j {\kappa}_k}{\Delta'})
 \right) -\nn\\
 && \frac{1}{3}
\left(\delta_{ij}+\frac{\bar{\kappa}_i \bar{\kappa}_j}{\Delta'} (1-\delta_{ij} {\kappa}_j) \right)  \,\,
 \left(\delta_{kl}+\frac{\kappa_k \kappa_l}{\Delta'} (1-\delta_{kl}\bar{\kappa}_l) \right) \q
 \ea
 where $\Delta'=\sum_i \kappa_i \bar{\kappa}_i$. Note that
\be
\kappa_i(1-\bar{\kappa}_i)= -\bar{\kappa}_i \q, \q\q \bar{\kappa}_i(1-\kappa_i)=-\kappa_i \q, \q\q (1-\kappa_i)(1-\bar{\kappa}_i)=1
\ee
so the  factors $(1-\delta_{ij}{\kappa}_j),(1-\delta_{ij}\bar{\kappa}_j)$  just change forward derivative into backward derivatives and vice versa.

 The projector $({}^1\!\Pi_{\rm L})$ projects onto the longitudinal tensor modes
\ba
v^k_{ij}=\delta^k_i \kappa_j (1 +(\delta^k_j-1)\kappa_i) + \delta^k_k k_i (1 +(\delta^k_i-1)k_j) \q , 
\ea
which constitute the gauge modes for the lattice version of the linearized Einstein-Hilbert Lagrangian (\ref{LEHL}), and represent the linearized diffeomorphism modes.

\begin{acknowledgments}
  BD thanks   Benjamin Knorr  and Jos\'e  Padua-Arg\"uelles   for interesting discussions. AK is supported by an NSERC grant awarded to BD.
Research at Perimeter Institute is supported in part by the Government of Canada through the Department of Innovation, Science and Economic Development Canada and by the Province of Ontario through the Ministry of Colleges and Universities.
\end{acknowledgments}

\vspace{1cm}

\begingroup
%\raggedright
\endgroup


\begin{thebibliography}{100}%\small

%\vspace{-2em}

\bibitem{BFCG2} S.~K.~Asante, B.~Dittrich, F.~Girelli, A.~Riello and P.~Tsimiklis,
  ``Quantum geometry from higher gauge theory,''
  arXiv:1908.05970 [gr-qc].
  %%CITATION = ARXIV:1908.05970;%%

\bibitem{EffSF1} S.~K.~Asante, B.~Dittrich and H.~M.~Haggard,
``Effective Spin Foam Models for Four-Dimensional Quantum Gravity,''
Phys. Rev. Lett. \textbf{125} (2020) no.23, 231301
%doi:10.1103/PhysRevLett.125.231301
[arXiv:2004.07013 [gr-qc]].

\bibitem{LQG} C. Rovelli,
\textit{Quantum Gravity},
(Cambridge University Press, Cambridge, 2004);
A. Ashtekar and J. Lewandowski,
``Background independent quantum gravity: a status report'',
Class. Quant. Grav. \textbf{21} R53 (2004), [arXiv:gr-qc/0404018];
T. ~Thiemann,
\textit{Introduction to Modern Canonical Quantum General Relativity},
(Cambridge University Press, Cambridge, 2007).
A.~Ashtekar and J.~Pullin,
\textit{Loop Quantum Gravity: The First 30 Years}, (World Scientific 2017)
%doi:10.1142/10445

\bibitem{Perez} A.~Perez,
  ``The Spin Foam Approach to Quantum Gravity,''
  Living Rev.\ Rel.\  {\bf 16} (2013) 3
 % doi:10.12942/lrr-2013-3
  [arXiv:1205.2019].
  %%CITATION = doi:10.12942/lrr-2013-3;%%

\bibitem{RyuTakayanagi} S.~Ryu and T.~Takayanagi, 
``Holographic Derivation of Entanglement Entropy from the anti--de Sitter Space/Conformal Field Theory Correspondence,''
Phys. Rev. Lett. {\bf 96} (2006) 181602.
%https://link.aps.org/doi/10.1103/PhysRevLett.96.181602

\bibitem{BHCounting} J.~D.~Bekenstein and V.~F.~Mukhanov,
  ``Spectroscopy of the quantum black hole,''
  Phys.\ Lett.\ B {\bf 360} (1995) 7
  %doi:10.1016/0370-2693(95)01148-J
  [gr-qc/9505012].
    A.~Ashtekar, J.~Baez, A.~Corichi and K.~Krasnov,
  ``Quantum geometry and black hole entropy,''
  Phys.\ Rev.\ Lett.\  {\bf 80} (1998) 904
 % doi:10.1103/PhysRevLett.80.904
  [gr-qc/9710007].
    J.~D.~Bekenstein,
  ``Statistics of black hole radiance and the horizon area spectrum,''
  Phys.\ Rev.\ D {\bf 91} (2015) no.12,  124052
 % doi:10.1103/PhysRevD.91.124052
  [arXiv:1505.03253 [gr-qc]].
  %%CITATION = doi:10.1103/PhysRevD.91.124052;%%
  J.~F.~Barbero G. and A.~Perez,
  ``Quantum Geometry and Black Holes,''
 % doi:10.1142/9789813220003_0008
  arXiv:1501.02963 [gr-qc].
  %%CITATION = doi:10.1142/9789813220003_0008;%%

\bibitem{Schuller1} F.~P.~Schuller and M.~N.~R.~Wohlfarth,
``Geometry of manifolds with area metric: multi-metric backgrounds,''
Nucl. Phys. B \textbf{747} (2006), 398-422
%doi:10.1016/j.nuclphysb.2006.04.019
[arXiv:hep-th/0508170 [hep-th]].

\bibitem{Schuller2} F.~P.~Schuller and M.~N.~R.~Wohlfarth,
``Canonical differential geometry of string backgrounds,''
JHEP \textbf{02} (2006), 059
%doi:10.1088/1126-6708/2006/02/059
[arXiv:hep-th/0511157 [hep-th]].
  R.~Punzi, F.~P.~Schuller and M.~N.~R.~Wohlfarth,
``Geometry for the accelerating universe,''
Phys. Rev. D \textbf{76} (2007), 101501
%doi:10.1103/PhysRevD.76.101501
[arXiv:hep-th/0612133 [hep-th]].
R.~Punzi, F.~P.~Schuller and M.~N.~R.~Wohlfarth,
``Area metric gravity and accelerating cosmology,''
JHEP \textbf{02} (2007), 030
%doi:10.1088/1126-6708/2007/02/030
[arXiv:hep-th/0612141 [hep-th]].

\bibitem{SFLimit} J.~W.~Barrett and R.~M.~Williams,
``The Asymptotics of an amplitude for the four simplex,''
Adv. Theor. Math. Phys. \textbf{3} (1999), 209-215
%doi:10.4310/ATMP.1999.v3.n2.a1
[arXiv:gr-qc/9809032 [gr-qc]].
F.~Conrady and L.~Freidel,
  ``On the semiclassical limit of 4d spin foam models,''
  Phys.\ Rev.\ D {\bf 78} (2008) 104023
  %doi:10.1103/PhysRevD.78.104023
  [arXiv:0809.2280 [gr-qc]].
  %%CITATION = doi:10.1103/PhysRevD.78.104023;%%
    J.~W.~Barrett, R.~J.~Dowdall, W.~J.~Fairbairn, H.~Gomes and F.~Hellmann,
  ``Asymptotic analysis of the EPRL four-simplex amplitude,''
  J.\ Math.\ Phys.\  {\bf 50} (2009) 112504
  doi:10.1063/1.3244218
  [arXiv:0902.1170 [gr-qc]];
  %%CITATION = doi:10.1063/1.3244218;%%
   J.~W.~Barrett, R.~J.~Dowdall, W.~J.~Fairbairn, F.~Hellmann and R.~Pereira,
  ``Lorentzian spin foam amplitudes: Graphical calculus and asymptotics,''
  Class.\ Quant.\ Grav.\  {\bf 27} (2010) 165009
  %doi:10.1088/0264-9381/27/16/165009
  [arXiv:0907.2440 [gr-qc]];
  %%CITATION = doi:10.1088/0264-9381/27/16/165009;%%
    M.~X.~Han and M.~Zhang,
  ``Asymptotics of Spinfoam Amplitude on Simplicial Manifold: Euclidean Theory,''
  Class.\ Quant.\ Grav.\  {\bf 29} (2012) 165004
 % doi:10.1088/0264-9381/29/16/165004
  [arXiv:1109.0500 [gr-qc]].
  %%CITATION = doi:10.1088/0264-9381/29/16/165004;%%


%

\bibitem{EffSF2} S.~K.~Asante, B.~Dittrich and H.~M.~Haggard,
``Discrete gravity dynamics from effective spin foams,''
Class. Quant. Grav. \textbf{38} (2021) no.14, 145023
%doi:10.1088/1361-6382/ac011b
[arXiv:2011.14468 [gr-qc]].

\bibitem{EffSF3} S.~K.~Asante, B.~Dittrich and J.~Padua-Arguelles,
``Effective spin foam models for Lorentzian quantum gravity,''
Class. Quant. Grav. \textbf{38} (2021) no.19, 195002
%doi:10.1088/1361-6382/ac1b44
[arXiv:2104.00485 [gr-qc]].

\bibitem{AreaRegge} C.~Rovelli,
``The Basis of the Ponzano-Regge-Turaev-Viro-Ooguri quantum gravity model in the loop representation basis,''
Phys. Rev. D \textbf{48} (1993), 2702-2707
%doi:10.1103/PhysRevD.48.2702
[arXiv:hep-th/9304164 [hep-th]].
J.~W.~Barrett, M.~Rocek and R.~M.~Williams,
  ``A Note on area variables in Regge calculus,''
  Class.\ Quant.\ Grav.\  {\bf 16} (1999) 1373
  %doi:10.1088/0264-9381/16/4/025
  [gr-qc/9710056];
  %%CITATION = doi:10.1088/0264-9381/16/4/025;%%

\bibitem{Regge} T.~Regge,
  ``General Relativity Without Coordinates,''
Nuovo Cim.  {\bf 19} (1961) 558.
  %doi:10.1007/BF02733251
  %%CITATION = doi:10.1007/BF02733251;%%

\bibitem{flatness} V.~Bonzom,
  ``Spin foam models for quantum gravity from lattice path integrals,''
  Phys.\ Rev.\ D {\bf 80} (2009) 064028
  %doi:10.1103/PhysRevD.80.064028
  [arXiv:0905.1501 [gr-qc]].
  %%CITATION = doi:10.1103/PhysRevD.80.064028;%%
   F.~Hellmann and W.~Kaminski,
  ``Holonomy spin foam models: Asymptotic geometry of the partition function,''
  JHEP {\bf 1310} (2013) 165
 % doi:10.1007/JHEP10(2013)165
  [arXiv:1307.1679 [gr-qc]].
  %%CITATION = doi:10.1007/JHEP10(2013)165;%%
    J.~R.~Oliveira,
  ``EPRL/FK Asymptotics and the Flatness Problem,''
  Class.\ Quant.\ Grav.\  {\bf 35} (2018) no.9,  095003
%  doi:10.1088/1361-6382/aaae82
  [arXiv:1704.04817 [gr-qc]].
   P.~Don\`a, F.~Gozzini and G.~Sarno,
  ``Searching for classical geometries in spin foam amplitudes: a numerical method,''
 % doi:10.1088/1361-6382/ab7ee1
  arXiv:1909.07832 [gr-qc].
  %%CITATION = doi:10.1088/1361-6382/ab7ee1;%%
J.~Engle, W.~Kaminski and J.~Oliveira,
``Addendum: EPRL/FK Asymptotics and the Flatness Problem,''
[arXiv:2012.14822 [gr-qc]].

\bibitem{MuxinSmallBI} M.~Han,
``On Spinfoam Models in Large Spin Regime,''
Class. Quant. Grav. \textbf{31} (2014), 015004
%doi:10.1088/0264-9381/31/1/015004
[arXiv:1304.5627 [gr-qc]].
M.~Han,
``Semiclassical Analysis of Spinfoam Model with a Small Barbero-Immirzi Parameter,''
Phys. Rev. D \textbf{88} (2013), 044051
doi:10.1103/PhysRevD.88.044051
[arXiv:1304.5628 [gr-qc]].

\bibitem{ComplexSP} M.~Han, Z.~Huang, H.~Liu and D.~Qu,
``Complex critical points and curved geometries in four-dimensional Lorentzian spinfoam quantum gravity,''
[arXiv:2110.10670 [gr-qc]].

\bibitem{DittrichRyan} B.~Dittrich and J.~P.~Ryan,
  ``Phase space descriptions for simplicial 4d geometries,''
  Class.\ Quant.\ Grav.\  {\bf 28} (2011) 065006
 % doi:10.1088/0264-9381/28/6/065006
  [arXiv:0807.2806 [gr-qc]];
  B.~Dittrich and J.~P.~Ryan,
  ``Simplicity in simplicial phase space,''
  Phys.\ Rev.\ D {\bf 82} (2010) 064026
 % doi:10.1103/PhysRevD.82.064026
  [arXiv:1006.4295 [gr-qc]].
  %%CITATION = doi:10.1103/PhysRevD.82.064026;%%
  ``On the role of the Barbero-Immirzi parameter in discrete quantum gravity,''
  Class.\ Quant.\ Grav.\  {\bf 30} (2013) 095015
%  doi:10.1088/0264-9381/30/9/095015
  [arXiv:1209.4892 [gr-qc]].

\bibitem{BC} J.~W.~Barrett and L.~Crane,
``Relativistic spin networks and quantum gravity,''
J. Math. Phys. \textbf{39} (1998), 3296-3302
%doi:10.1063/1.532254
[arXiv:gr-qc/9709028 [gr-qc]].

\bibitem{EPRL-FK} J.~Engle, R.~Pereira and C.~Rovelli,
  ``The Loop-quantum-gravity vertex-amplitude,''
  Phys.\ Rev.\ Lett.\  {\bf 99} (2007) 161301
 % doi:10.1103/PhysRevLett.99.161301
  [arXiv:0705.2388 [gr-qc]];
  %%CITATION = doi:10.1103/PhysRevLett.99.161301;%%
  L.~Freidel and K.~Krasnov,
  ``A New Spin Foam Model for 4d Gravity,''
  Class.\ Quant.\ Grav.\  {\bf 25} (2008) 125018
  %doi:10.1088/0264-9381/25/12/125018
  [arXiv:0708.1595 [gr-qc]];
       E.~R.~Livine and S.~Speziale,
  ``Consistently Solving the Simplicity Constraints for Spinfoam Quantum Gravity,''
  EPL {\bf 81} (2008) no.5,  50004
 % doi:10.1209/0295-5075/81/50004
  [arXiv:0708.1915 [gr-qc]];
  %%CITATION = doi:10.1088/0264-9381/25/12/125018;%%
    J.~Engle, E.~Livine, R.~Pereira and C.~Rovelli,
  ``LQG vertex with finite Immirzi parameter,''
  Nucl.\ Phys.\ B {\bf 799} (2008) 136
  %doi:10.1016/j.nuclphysb.2008.02.018
  [arXiv:0711.0146 [gr-qc]];
  %%CITATION = doi:10.1016/j.nuclphysb.2008.02.018;%%
  A.~Baratin and D.~Oriti,
``Group field theory and simplicial gravity path integrals: A model for Holst-Plebanski gravity,''
Phys. Rev. D \textbf{85} (2012), 044003
%doi:10.1103/PhysRevD.85.044003
[arXiv:1111.5842 [hep-th]].

\bibitem{ARE1} B.~Dittrich,
``Modified Graviton Dynamics From Spin Foams: The Area Regge Action,''
[arXiv:2105.10808 [gr-qc]].

\bibitem{ReggeWilliams} T.~Regge and R.~M.~Williams,
``Discrete structures in gravity,''
J. Math. Phys. \textbf{41} (2000), 3964-3984
%doi:10.1063/1.533333
[arXiv:gr-qc/0012035 [gr-qc]].

\bibitem{LinkNB} {\href{https://github.com/star-pirate/Centrally_Subdivided_Area_Regge}{ Link to GitHub Repository}}

\bibitem{RocekWilliams} M.~Rocek and R.~M.~Williams,
``Quantum Regge Calculus,''
Phys. Lett. B \textbf{104} (1981), 31.
%doi:10.1016/0370-2693(81)90848-0
M.~Rocek and R.~M.~Williams,
``The Quantization of Regge Calculus,''
Z. Phys. C \textbf{21} (1984), 371 .
%doi:10.1007/BF01581603

\bibitem{Hol3D4D} V.~Bonzom and B.~Dittrich,
``3D holography: from discretum to continuum,''
JHEP \textbf{03} (2016), 208
%doi:10.1007/JHEP03(2016)208
[arXiv:1511.05441 [hep-th]].
S.~K.~Asante, B.~Dittrich and H.~M.~Haggard,
``Holographic description of boundary gravitons in (3+1) dimensions,''
JHEP \textbf{01} (2019), 144
%doi:10.1007/JHEP01(2019)144
[arXiv:1811.11744 [hep-th]].

\bibitem{PImeasure1} B.~Dittrich and S.~Steinhaus,
  ``Path integral measure and triangulation independence in discrete gravity,''
  Phys.\ Rev.\ D {\bf 85} (2012) 044032
 % doi:10.1103/PhysRevD.85.044032
  [arXiv:1110.6866 [gr-qc]];
  %%CITATION = doi:10.1103/PhysRevD.85.044032;%%
  B.~Dittrich, W.~Kami\'nski and S.~Steinhaus,
``Discretization independence implies non-locality in 4D discrete quantum gravity,''
Class. Quant. Grav. \textbf{31} (2014) no.24, 245009
%doi:10.1088/0264-9381/31/24/245009
[arXiv:1404.5288 [gr-qc]].
S.~K.~Asante and B.~Dittrich,
``Perfect discretizations as a gateway to one-loop partition functions for 4D gravity,''
[arXiv:2112.03307 [gr-qc]].

\bibitem{Improved} B.~Bahr and B.~Dittrich,
``Improved and Perfect Actions in Discrete Gravity,''
Phys. Rev. D \textbf{80} (2009), 124030
%doi:10.1103/PhysRevD.80.124030
[arXiv:0907.4323 [gr-qc]].

\bibitem{BahrDittrichHe} B.~Bahr, B.~Dittrich and S.~He,
``Coarse graining free theories with gauge symmetries: the linearized case,''
New J. Phys. \textbf{13} (2011), 045009
%doi:10.1088/1367-2630/13/4/045009
[arXiv:1011.3667 [gr-qc]].





%

\bibitem{ContLimit} B.~Dittrich,
``The continuum limit of loop quantum gravity - a framework for solving the theory,''
doi:10.1142/9789813220003\_0006
[arXiv:1409.1450 [gr-qc]].
B.~Dittrich, F.~C.~Eckert and M.~Martin-Benito,
``Coarse graining methods for spin net and spin foam models,''
New J. Phys. \textbf{14} (2012), 035008
%doi:10.1088/1367-2630/14/3/035008
[arXiv:1109.4927 [gr-qc]].
C.~Delcamp and B.~Dittrich,
``Towards a phase diagram for spin foams,''
Class. Quant. Grav. \textbf{34} (2017) no.22, 225006
%doi:10.1088/1361-6382/aa8f24
[arXiv:1612.04506 [gr-qc]].
 S.~Steinhaus,
``Coarse Graining Spin Foam Quantum Gravity\textemdash{}A Review,''
Front. in Phys. \textbf{8} (2020), 295
%doi:10.3389/fphy.2020.00295
[arXiv:2007.01315 [gr-qc]].

\bibitem{ADH1} S.~K.~Asante, B.~Dittrich and H.~M.~Haggard,
  ``The Degrees of Freedom of Area Regge Calculus: Dynamics, Non-metricity, and Broken Diffeomorphisms,''
  Class.\ Quant.\ Grav.\  {\bf 35} (2018) no.13,  135009
 % doi:10.1088/1361-6382/aac588
  [arXiv:1802.09551]
  %%CITATION = doi:10.1088/1361-6382/aac588;%%

\bibitem{AreaAngle} B.~Dittrich and S.~Speziale,
  ``Area-angle variables for general relativity,''
  New J.\ Phys.\  {\bf 10} (2008) 083006
%  doi:10.1088/1367-2630/10/8/083006
  [arXiv:0802.0864 [gr-qc]].
  %%CITATION = doi:10.1088/1367-2630/10/8/083006;%%

\bibitem{DittrichHoehn1} B.~Dittrich and P.~A.~H\"ohn,
  ``Canonical simplicial gravity,''
  Class. Quant. Grav.  {\bf 29} (2012) 115009
 % doi:10.1088/0264-9381/29/11/115009
  [arXiv:1108.1974 [gr-qc]].
  %%CITATION = doi:10.1088/0264-9381/29/11/115009;%%
B.~Dittrich and P.~A.~H\"ohn,
``Constraint analysis for variational discrete systems,''
J. Math. Phys. \textbf{54} (2013), 093505
%doi:10.1063/1.4818895
[arXiv:1303.4294 [math-ph]].

\bibitem{Mara} P.~S.~Mara, ``Triangulations for the cube" Journal of Combinatorical Theory (A) {\bf 20} (1976) 170

\bibitem{DittrichFreidelSpeziale} B.~Dittrich, L.~Freidel and S.~Speziale,
``Linearized dynamics from the 4-simplex Regge action,''
Phys. Rev. D \textbf{76} (2007), 104020
%doi:10.1103/PhysRevD.76.104020
[arXiv:0707.4513 [gr-qc]].

\bibitem{DiffReview08} B.~Dittrich,
``Diffeomorphism symmetry in quantum gravity models,''
Adv. Sci. Lett. \textbf{2}, 151
%doi:10.1166/asl.2009.1022
[arXiv:0810.3594 [gr-qc]].

\bibitem{BarrettWilliams} J.~W.~Barrett,
``The fundamental theorem of linearized Regge calculus,''
Phys. Lett. B \textbf{190} (1987), 135-136.
%doi:10.1016/0370-2693(87)90853-7
J.~W.~Barrett and R.~M.~Williams,
``The Convergence of Lattice Solutions of Linearized Regge Calculus,''
Class. Quant. Grav. \textbf{5} (1988), 1543-1556.
%doi:10.1088/0264-9381/5/12/007

\bibitem{Wainwright} C.~Wainwright and R.~M.~Williams,
``Area Regge calculus and discontinuous metrics,''
Class. Quant. Grav. \textbf{21} (2004), 4865-4880
%doi:10.1088/0264-9381/21/21/008
[arXiv:gr-qc/0405031 [gr-qc]].

\bibitem{HamberWilliamsHO} H.~W.~Hamber and R.~M.~Williams,
``Simplicial Quantum Gravity With Higher Derivative Terms: Formalism and Numerical Results in Four-dimensions,''
Nucl. Phys. B \textbf{269} (1986), 712-743
%doi:10.1016/0550-3213(86)90518-3

\bibitem{PerfectPI} B.~Bahr, B.~Dittrich and S.~Steinhaus,
``Perfect discretization of reparametrization invariant path integrals,''
Phys. Rev. D \textbf{83} (2011), 105026
doi:10.1103/PhysRevD.83.105026
[arXiv:1101.4775 [gr-qc]].

\bibitem{Krasnov} K.~Krasnov,
  ``Gravity as BF theory plus potential,''
  Int.\ J.\ Mod.\ Phys.\ A {\bf 24} (2009) 2776
  %doi:10.1142/S0217751X09046151
  [arXiv:0907.4064 [gr-qc]];
  %%CITATION = doi:10.1142/S0217751X09046151;%%
    K.~Krasnov,
  ``Effective metric Lagrangians from an underlying theory with two propagating degrees of freedom,''
  Phys.\ Rev.\ D {\bf 81} (2010) 084026
 % doi:10.1103/PhysRevD.81.084026
  [arXiv:0911.4903 [hep-th]].
  %%CITATION = doi:10.1103/PhysRevD.81.084026;%%

\bibitem{Alesci} E.~Alesci and C.~Rovelli,
``The Complete LQG propagator. I. Difficulties with the Barrett-Crane vertex,''
Phys. Rev. D \textbf{76} (2007), 104012
%doi:10.1103/PhysRevD.76.104012
[arXiv:0708.0883 [gr-qc]].

\bibitem{DG} B.~Dittrich and M.~Geiller,
``A new vacuum for Loop Quantum Gravity,''
Class. Quant. Grav. \textbf{32} (2015) no.11, 112001
%doi:10.1088/0264-9381/32/11/112001
[arXiv:1401.6441 [gr-qc]].
B.~Bahr, B.~Dittrich and M.~Geiller,
``A new realization of quantum geometry,''
Class. Quant. Grav. \textbf{38} (2021) no.14, 145021
%doi:10.1088/1361-6382/abfed1
[arXiv:1506.08571 [gr-qc]].

\bibitem{SFDiv} C.~Perini, C.~Rovelli and S.~Speziale,
``Self-energy and vertex radiative corrections in LQG,''
Phys. Lett. B \textbf{682} (2009), 78-84
%doi:10.1016/j.physletb.2009.10.076
[arXiv:0810.1714 [gr-qc]].
A.~Riello,
``Self-energy of the Lorentzian Engle-Pereira-Rovelli-Livine and Freidel-Krasnov model of quantum gravity,''
Phys. Rev. D \textbf{88} (2013) no.2, 024011
%doi:10.1103/PhysRevD.88.024011
[arXiv:1302.1781 [gr-qc]].
V.~Bonzom and B.~Dittrich,
``Bubble divergences and gauge symmetries in spin foams,''
Phys. Rev. D \textbf{88} (2013), 124021
%doi:10.1103/PhysRevD.88.124021
[arXiv:1304.6632 [gr-qc]].
L.~Q.~Chen,
``Bulk amplitude and degree of divergence in 4d spin foams,''
Phys. Rev. D \textbf{94} (2016) no.10, 104025
%doi:10.1103/PhysRevD.94.104025
[arXiv:1602.01825 [gr-qc]].

\bibitem{DiffDiv} B.~Bahr and B.~Dittrich,
``(Broken) Gauge Symmetries and Constraints in Regge Calculus,''
Class. Quant. Grav. \textbf{26} (2009), 225011
%doi:10.1088/0264-9381/26/22/225011
[arXiv:0905.1670 [gr-qc]].
B.~Bahr and B.~Dittrich,
``Breaking and restoring of diffeomorphism symmetry in discrete gravity,''
AIP Conf. Proc. \textbf{1196} (2009) no.1, 10
%doi:10.1063/1.3284371
[arXiv:0909.5688 [gr-qc]].

\bibitem{BahrStein} B.~Bahr and S.~Steinhaus,
``Numerical evidence for a phase transition in 4d spin foam quantum gravity,''
Phys. Rev. Lett. \textbf{117} (2016) no.14, 141302
%doi:10.1103/PhysRevLett.117.141302
[arXiv:1605.07649 [gr-qc]].

\bibitem{Ambjorn:1997ub} J.~Ambjorn, J.~L.~Nielsen, J.~Rolf and G.~K.~Savvidy,
%``Spikes in quantum Regge calculus,''
Class. Quant. Grav. \textbf{14} (1997), 3225-3241
doi:10.1088/0264-9381/14/12/009
[arXiv:gr-qc/9704079 [gr-qc]].

\bibitem{Gibbons:1978ac} G.~W.~Gibbons, S.~W.~Hawking and M.~J.~Perry,
``Path Integrals and the Indefiniteness of the Gravitational Action,''
Nucl. Phys. B \textbf{138} (1978), 141-150

\bibitem{LollLR} R.~Loll,
``Discrete approaches to quantum gravity in four-dimensions,''
Living Rev. Rel. \textbf{1} (1998), 13
%doi:10.12942/lrr-1998-13

\bibitem{LorCal} J.~Feldbrugge, J.~L.~Lehners and N.~Turok,
``Lorentzian Quantum Cosmology,''
Phys. Rev. D \textbf{95} (2017) no.10, 103508
%doi:10.1103/PhysRevD.95.103508
[arXiv:1703.02076 [hep-th]].
M.~Han, Z.~Huang, H.~Liu, D.~Qu and Y.~Wan,
``Spinfoam on a Lefschetz thimble: Markov chain Monte Carlo computation of a Lorentzian spinfoam propagator,''
Phys. Rev. D \textbf{103} (2021) no.8, 084026
%doi:10.1103/PhysRevD.103.084026
[arXiv:2012.11515 [gr-qc]].
D.~Jia,
``Complex, Lorentzian, and Euclidean simplicial quantum gravity: numerical methods and physical prospects,''
[arXiv:2110.05953 [gr-qc]].
S.~K.~Asante, B.~Dittrich and J.~Padua-Arg\"uelles,
``Complex actions and causality violations: Applications to Lorentzian quantum cosmology,''
[arXiv:2112.15387 [gr-qc]].


\end{thebibliography}
\end{document}